\documentclass{aa}
\usepackage{txfonts}
\usepackage{graphicx}
\usepackage{natbib}
\bibpunct{(}{)}{;}{a}{}{,}

\begin{document}

\title{On the dynamics and collisional growth of planetesimals 
in misaligned binary systems}

\author{M.M. Fragner \inst{1}, R.P. Nelson \inst{1}  \& W. Kley \inst{2}}
\institute{Astronomy Unit, Queen Mary, University of London, Mile
End Road, London E1 4NS, U.K. 
\and
Institut fur Astronomie \& Astrophysik, Universitat Tubingen, Auf der Morgenstelle 10, 
72076 Tubingen, Germany \\
\email{M.Fragner@qmul.ac.uk, R.P.Nelson@qmul.ac.uk, wilhelm.kley@uni-tubingen.de}}
\date{Received/Accepted}

\abstract
{Abridged. Many stars are members of binary systems. During early phases
when the stars are surrounded by
a discs, the binary orbit and
disc midplane may be mutually inclined. The
discs around T Tauri stars 
will become mildly warped and undergo
solid body precession around the angular momentum vector of
the binary system. It is unclear how
planetesimals embedded in such a disc will evolve
and affect planet formation.}
{We investigate the dynamics of planetesimals embedded in gaseous
protoplanetary discs that are perturbed by a binary
companion on a circular, inclined orbit. 
We examine the collisional velocities of the planetesimals
to determine when planetesimal growth
through accretion
can occur, instead of disruption.
We vary the binary inclination, $\gamma_F$, 
binary separation, $D$, disc mass, $M_d$, and
planetesimal radius $s_i$. Our standard model has $D=60$ AU,
$\gamma_F=45^{\circ}$, 
and a disc mass equivalent to that of the minimum mass solar nebula model.}
{We use a 3D hydrodynamics code to model the  
disc.
Planetesimals are test particles which
experience gas drag, the gravitational force of the disc, 
the companion star gravity.
Planetesimal orbital crossing events are detected and used to
estimate collisional velocities.}
{For binary systems with modest inclination ($\gamma_F=25^\circ$),
disc gravity prevents planetesimal orbits
from undergoing strong differential nodal precession
(which they would do in the absence of the disc), and forces
the planetesimals to precess with the disc on average. 
For planetesimals of different size the orbit planes
become modestly mutually inclined, leading to
collisional velocities that inhibit planetesimal growth.
For larger inclinations ($\gamma_F=45^\circ$), the
Kozai effect operates, leading to destructively large 
relative velocities.}
{We conclude that planet formation via planetesimal accretion
is difficult in an inclined binary system with
parameters similar to those considered in this paper.
For highly inclined systems in which
the Kozai effect switches on, the prospects for forming planets
are very remote indeed.}

\keywords{}
\titlerunning{On collisional growth of planetesimals in misaligned binary systems}
\authorrunning{M.M. Fragner, R.P. Nelson \& W. Kley}

\maketitle

\section{Introduction}
Of the extrasolar planets detected so far, over 20 percent are found to orbit
one component of a multiple/binary star system \citep{desidera, eggenberger}.
Planet formation in binary systems can represent a particular challenge,
as each stage of the formation process can be affected 
by the gravitational perturbation of the binary companion. 
One crucial stage that is particularly sensitive to the presence of the 
companion star is the accumulation stage of kilometre-sized planetesimals.

The fundamental parameter that controls the efficiency of planetary growth by accretion
of planetesimals is the relative velocity between impacting objects.
This must be lower than the
escape velocity from the accreting objects in order for efficient runaway
accretion to occur \citep{wetherill}. If the external gravitational
perturbation by the binary companion excites relative velocities 
that exceed the escape velocity, runway
accretion is prevented and growth remains slow. 
Furthermore, if the relative velocity is excited beyond
the threshold velocity at which erosion dominates accretion, 
planetesimal growth potentially ceases altogether \citep{benz}.

The possible effect of a stellar companion on the perturbed
velocity distribution of planetesimals has been explored in 
several previous publications, where most have
considered configurations in which the binary orbit is eccentric and
coplanar with the disc midplane \citep{heppenheimer, marzari, thebault, paardekooper1}.
The main conclusion of these studies was that the coupled effects of secular 
perturbations of the binary companion, and friction due to gas drag arising from
the protoplanetary gas disc, induce forced eccentricities and
a size-dependent phasing of pericentres. This leads to relatively
modest collision velocities dominated by the keplerian shear for same-sized bodies,
but high relative velocities for different sized planetesimals.
As a consequence, binary systems with separations $\sim 50$ AU may have a 
strong inhibiting effect on accretion within a swarm of colliding km-size
planetesimals.

These studies neglected the effect of disc gravity acting on the planetesimals,
which can affect the details of the results when the disc becomes eccentric 
through interaction with a binary system on an eccentric orbit. But, the
main conclusion that planetesimals of different sizes experience large
collisional velocities due to differential phasing of
their pericentres remains valid \citep{kley}.

These previous studies assumed that the orbital plane of the binary companion 
is coplanar with the disc midplane.
According to \cite{hale} the assumption of coplanarity or modest inclination
may be reasonable
for binary separations below $\sim 40$ AU, but systems with larger separations 
appear to have their orbital planes randomly distributed.
The examination of planetesimal dynamics in non-coplanar configurations 
has received attention only recently \citep{marzari2, xie}.
The former authors show that, due to the semi-major axis dependence of the 
nodal precession rate, the nodal lines of the planetesimals become progressively 
randomized. This may lead to the dispersion of the planetesimal disk that expands
into a cloud of bodies surrounding the star. The latter authors considered
the effect of gas drag in an system with a modestly inclined binary, and showed that
in addition to inducing size-dependent phasing of pericentres, gas drag also
introduces phasing of nodal lines. This leads to a situation is which
planetesimals of different size occupy different orbital planes.
\citet{xie} suggest that this may provide a favourable channel for
planetesimal growth as low velocity collisions between similar sized objects
become more frequent than high velocity collisions between different sized
objects.
It should be noted, however, that these authors neglected the effect of disc 
gravity on the planetesimals. Although the inclusion of disc gravity
modifies the dynamics of planetesimals in fully coplanar systems, the
changes are not dramatic. In non-coplanar systems, however, where
the disc and planetesimals may precess at different rates around the
binary angular momentum vector, the deviation of the planetesimals
from the disc plane results in the disc gravity providing a strong 
restoring force that can modify the dynamics in an important qualitative sense.

The recent simulations by \cite{marzari2} and \cite{xie} either ignored
the gaseous protoplanetary disc, or treated it as a static object which did not 
evolve in the gravitational field of the inclined companion.
It is known, however, that a gaseous disc orbiting under the influence of
a close binary companion on an inclined orbit will develop a mild warp
and precess as a solid body around the orbital angular momentum vector of the
binary system if bending waves can propagate efficiently \citep{pringle,
paplin, ogilvie, papterquem, lubow}.
This process has been studied using numerical simulations by \cite{larwood}, who
used Smooth Particle Hydrodynamic calculations, and more recently
by \cite{fragner}, who performed 3D simulations using a grid-based
hydrodynamics code . 
The condition for bending waves to propagate is that
$\alpha < H/R$, where $\alpha$ is the usual viscosity parameter \citep{shakura}
and $H/R$ is the disc aspect ratio. In protostellar discs, it is estimated
$H/R \simeq 0.05$ and $\alpha \simeq 10^{-3}$ - $10^{-2}$, so we expect such discs
to evolve similarly to the models presented in this paper. In addition to
generating a mild warp, and causing the disc to precess, the binary can
also tidally truncate the disc at its outer edge, where the tidal
truncation radius of the disc for a binary system of unit mass
ratio is typically $\sim D/3$, where $D$ is the
binary separation \citep{artymowicz, larwood, fragner}.

An observational example of a disc in a misaligned binary system
is HK Tau \citep{stapelfeldt}, where the binary system and disc have been 
imaged directly. More indirect evidence for precessing discs in
close binary systems comes from observations of precessing
jets in star forming regions \citep{Terquem}.

In this paper we investigate the dynamics of planetesimals embedded in
protoplanetary disc models in inclined binary systems with circular orbits.
Although most binaries are on eccentric orbits, we chose the case of a 
circular orbit to allow us to focus on the effects inclination. In 
contrast to previous work, we solve for the disc evolution in the 
gravitational field of the inclined binary companion, and include the
effects of the disc gravity acting on the planetesimals. 
We consider two different values of the inclination between the binary 
orbit plane and disc midplane, $\gamma_F=25^\circ$ and $\gamma_F=45^\circ$,
planetesimals of size 100 m, 1 km and 10 km, and binary separations
between 60 and 120 AU.  The disc outer radius is 18 AU, consistent with
the tidal truncation radius for a binary separation $D=60$ AU.

Due to the complexity of our model, and 
the associated computational
expense of running the simulations, we are unable to consider large
numbers of planetesimals, and so we are forced to use an approximate
method for determining their collisional velocities based on examining
the moments when the osculating orbits of the planetesimals intersect.
Using this approximation, we find that in general the binary companion
causes large planetesimal collision velocities to be generated,
largely because the orbit planes of the planetesimals become mutually
inclined. For the larger value of the inclination, we find that the
Kozai mechanism can switch on, leading to the generation of
large orbital eccentricities for the planetesimals, and therefore
very large collision velocities. These results indicate that planet
formation via the accumulation of planetesimals will be difficult
in binary systems with parameters similar to those we consider in this paper.

This paper is organised as follows. In Sect.~2 we present the basic 
equations and in Sect.~3 we
discuss the numerical techniques. In Sect.~4 we investigate numerically the 
effect of disc gravity on
the evolution of inclined planetesimal orbits in the absence of a binary
companion. In Sect.~5 we examine the dynamics of planetesimals that
are embedded in a disc which is inclined initially with respect to 
the binary plane, and examine the effect of varying the inclination angle 
$\gamma_F$. 
In section Sect.~6 we present calculations
for different disc masses and binary separations and 
examine under which condition the  Kozai effect can be suppressed. 
Finally, we discuss our results and draw conclusions in Sect.~7.

\section{Basic equations}
\label{basic}
The equations of hydrodynamics for a viscous disc that we solve in
this paper are given in \cite{fragner}.
The disc evolves under the effects of 
pressure, viscosity and the gravitational forces due to
the binary companion and central star. We employ a large number of variables
in this paper to describe the results, and we tabulate these in
table~\ref{table1} for ease of reference.

We solve the equations in a frame that precesses around the angular 
momentum vector of the binary orbit, since it is expected that
the disc models we consider in this paper will undergo uniform
precession due to interaction with the binary companion. In this frame
the disc midplane stays close to the equatorial plane of the
spherical polar grid that we adopted in the simulations, provided that 
the precession frequency of the frame, 
$\Omega_F$, is chosen according to \citep{fragner}:
\begin{eqnarray}
\frac{\Omega_F}{\Omega_d(R)}=-\frac{3}{7}\frac{R^3}{D^3}
\frac{M_B}{M_\star} \cos{(\gamma_F)},
\label{omega}
\end{eqnarray}
where $\Omega_d(R)$ is the orbital angular velocity of
the disc at its outer radius $R$, $D$ is the binary separation, 
$M_\star$ and $M_B$ are the masses of the primary and secondary star, 
respectively, and $\gamma_F$ is the relative inclination between the 
disc midplane and the binary orbital plane.

The binary companion is held on a fixed circular orbit with separation $D$.
Its position vector, ${\bf D}$, in the precessing frame is given by \citep{fragner}:
\begin{eqnarray}
{\bf D}=D\left(\begin{array}{c}\cos{([\omega_B-\Omega_F]t)}\\
\sin{([\omega_B - \Omega_F]t)} \cos(\gamma_F)\\
\sin{([\omega_B-\Omega_F]t))} \sin{(\gamma_F)}
\end{array} \right)
\end{eqnarray}
where $\omega_B$ is the binary angular frequency measured in the 
non-precessing binary frame.
Thus an observer moving with the precessing frame sees an 
increased binary frequency
$\omega_B -\Omega_F$ due to the retrograde precession of the frame
(i.e. $\Omega_F$ is negative).

The planetesimals we model do not mutually interact. They feel
the gravitational force of the primary star, secondary star and disc, 
the drag force due to the disc, Coriolis
and centrifugal forces due to the precession of the frame and 
indirect forces that arise because we centre
our coordinate system on the primary star. The equations describing 
the evolution of the planetesimals are therefore:
\begin{eqnarray}
\frac{\partial^2{\bf r}_i}{\partial t^2}=&-&\frac{GM_\star}{r_i^3}{\bf r}_i-GM_B\frac{({\bf r}_i-{\bf D})}{|{\bf r}_i-{\bf D}|^3}-G\int_{disc}dm({\bf r})\frac{({\bf r}_i-{\bf r})}{|{\bf r}_i-{\bf r}|^3}\nonumber\\
&+&\frac{{\bf F}_D}{M_i}-2{\bf\Omega}_F\times{\bf v_i}-{\bf\Omega}_F\times ({\bf\Omega}_F\times{\bf r_i})\nonumber\\
&-&\frac{GM_B}{D^3}{\bf D}-G\int_{disc}dm({\bf r})\frac{{\bf r}}{r^3}
\end{eqnarray}
where $G$ is the gravitational constant, ${\bf r}_i$, 
${\bf v}_i$ and $M_i$ are the position and velocity vector
and mass of planetesimal $i$. The first and second terms are the
gravitational acceleration
due to the central and companion star, respectively, and the 
third represents the gravitational acceleration due to the disc.
The fifth and sixth terms represent Coriolis and centrifugal 
forces that arise because the
coordinate system precesses around a vector ${\bf\Omega}_F$, 
given by \citep{fragner}:
\begin{eqnarray}
{\bf\Omega}_F=\Omega_F\left(\begin{array}{c}0\\
-\sin(\gamma_F)\\
\cos{(\gamma_F)}
\end{array} \right).
\end{eqnarray}
The last two terms are
indirect terms that account for the acceleration of 
the central star by the companion and disc, respectively. The
gas drag force ${\bf F}_D$ can be written \citep{weidenschilling}:
\begin{eqnarray}
{\bf F}_D=-C_D\pi s_i^2\rho\frac{1}{2}|{\bf v}_i-{\bf v}|({\bf v}_i-{\bf v})
\label{dragforce}
\end{eqnarray}
Here $s_i$ is the radius of planetesimal $i$, $\rho$ and ${\bf v}$ are the gas disc density and velocity.
The value of the drag coefficient $C_D$ depends on the Reynolds number
\begin{equation}
{\cal R}_e=\frac{2s_i|{\bf v}_i-{\bf v}|}{\nu_{mol}}
\end{equation}
where the molecular viscosity, $\nu_{mol}=\lambda_M c_s$,
$\lambda_M$ is the mean free path of the gas molecules,
and $c_s$ is the sound speed.
$C_D$ takes values:
\begin{eqnarray}
   & & \ 24. \ {\cal R}_e^{-1}  \;\;\;\;\;\;\; {{\cal R}_e < 1}  \nonumber \\
  C_D & = & \ 24.\ {\cal R}_e^{-0.6}  \;\;\;\;\; {1 < {\cal R}_e \le 800}  \nonumber \\
 & &  \ 0.44 \; \;\;\;\;\;\;\;\;\;\; {{\cal R}_e > 800}  
\label{eqn:C_D}
\end{eqnarray}

\section{Numerical method}
The hydrodynamic disc equations
are integrated using the grid-based hydrodynamics
code NIRVANA \citep{ziegler}, adapted to solve the equations
in a precessing reference frame. This code uses operator-splitting,
and the advection routine uses a second-order
accurate monotonic transport algorithm \citep{leer}.
The planetesimal orbits are integrated using the leap-frog integrator.

\subsection{Units}
The equations are integrated in dimensionless form, where we choose 
our unit of length to be the
radius of the disc inner edge. The unit of mass is that of 
the central star. The unit of time
used in the code corresponds to $\Omega_K(1)^{-1}$
(where $\Omega_K(1)$ is the keplerian angular frequency at radius $R=1$),
and the gravitational constant $G=1$.
When discussing simulation results we will refer to a time unit that corresponds to
$P_d=2\pi/\Omega_K(10)$, which is approximately one orbit at the outer edge 
of the disc. In the sections of this paper which describe the results of
simulations we refer to evolution times in units of ``orbits'', where an
orbit should be understood as being a time interval equal to $P_d$.
We scale our unit of length to $2$ AU in physical units, and
assume that the central star has one solar mass,
so that one
orbital period at the outer disc edge corresponds to a time
of $P_d=89.44$ years.
Inclination and precession angles are displayed in units of radians.

\subsection{Initial and boundary conditions}
The disc model we use is model 1 described in \cite{fragner},
and interested readers are referred  to that paper for a general
description of disc evolution in misaligned binary systems.
The disc model extends from $1-9$ in code units,
corresponding to $2-18$ AU in physical units, has a height to radius ratio 
$h=0.05$, and a dimensionless viscosity parameter $\alpha=0.025$
\citep{shakura}.
During its evolution, the disc model develops a very modest warp
(the variation in inclination across its radial extent is less than
one degree), and it precesses as a
rigid body around the angular momentum vector of the binary system 
with a frequency approximately equal to $\Omega_F$ given by Eq.~(\ref{omega}).
The disc exhibits spiral density waves which are excited by the companion,
and does not appear to become noticeably
eccentric during its evolution -- unlike discs which
are perturbed by lower mass binary companions \citep{kley-papaloizou},
or companions on eccentric orbits \citep{kley}.

In our standard model we normalise the disc mass so
that it would contain 0.015 M$_{\odot}$ if extended out to a radius of 40 AU
(this being very similar to the mass contained in the
minimum mass solar nebula model \citep{hayashi}), although
the model we use nominally only goes out to $\sim 18$ AU.
The actual disc mass is $M_d=\hat{M}= 4.52 \times 10^{-3}M_\odot$, 
where $\hat{M}$ denotes our standard disc mass.
In physical units the gas disc density is given by:
$$\rho(r,\hat{\theta})=7.1\cdot 10^{-11}\left(\frac{r}{\rm AU}\right)^{-\frac{3}{2}} 
\sin{(\hat{\theta}})^{\left(\frac{1}{h^2}-\frac{5}{2}\right)} \;\; \rm{g} \;  {\rm cm}^{-3}, $$
where $\hat \theta$ is the meridional angle measured relative to
the normal to the disc midplane.
The binary companion is held on a circular inclined orbit with constant 
separation $D=30$ ($\equiv 60$ AU).
We note that most binaries are
on eccentric orbits \citep{duquennoy}, but we restrict ourselves
to circular inclined orbits in this study so as to isolate effects
relating to the inclination.
Note that we neglect the disc gravity acting on the binary companion,
so the binary orbit is fixed.
At the beginning of each simulation, the companion mass is 
increased linearly until it
reaches its final mass $M_B=1M_\star$ after 4 orbits. 
Its angular frequency $\omega_B$ is
increased accordingly to be consistent with a stationary orbit.

Periodic boundary conditions were applied in the azimuthal direction. At all 
other boundaries standard stress-free, outflow conditions were employed.

\subsection{Planetesimal set-up}
The mass of the planetesimals is given by $M_i=\frac{4}{3}\pi s_i^3\rho_s$, 
and we choose a solid density of
$\rho_s=2$ g cm$^{-3}$. They are initially embedded within the 
unperturbed disc model on circular, keplerian orbits 
which are coplanar with the disc midplane. As the disc does not become
noticeably warped or eccentric during its evolution, it is not necessary
to pre-evolve the disc so that it achieves a steady state structure prior
to inserting the planetesimals. 

In the following discussion we will characterise the planetesimal evolution
using their orbital elements in the fixed binary frame,
and these are calculated from the positions and velocities
that the code outputs in the precessing frame.
Hence, we first have to transform the velocities and positions 
from the precessing frame into the binary frame, in which the
angular momentum vector of the binary 
${\bf J}_B$ is parallel to the unit vector 
${\bf e}_3$. 

As we consider two reference frames in this paper, one which 
precesses with the disc, and the fixed frame based on the
binary system, we denote vectors and coordinate values
in the precessing frame using the hat-symbol 
(e.g. $\hat {\bf r}$, $\hat \theta$). Vectors and coordinate
values in the fixed frame are denoted without the hat-symbol
(e.g. ${\bf r}$, $\theta$).
The transformation of the 
position and velocity vectors ${\bf\hat{r}}_i$ and  
${\bf\hat{v}}_i$ defined in the precessing frame into 
vectors ${\bf r}_i$ and ${\bf v}_i$ defined in the 
non-precessing binary frame is given by \citep{fragner}:
\begin{eqnarray}
{\bf r}_i&=&R_X^{-1}(\gamma_F)R_Z^{-1}(\Omega_Ft){\bf\hat r}_i\nonumber\\
{\bf v}_i&=&\frac{\partial}{\partial t}{\bf r}_i=R_X^{-1}(\gamma_F)R_Z^{-1}(\Omega_Ft){\bf\hat v}_i+{\bf\Omega}_F\times{\bf r}_i
\label{trafo}
\end{eqnarray}
where $R_Z$ and $R_X$ are rotation matrices around the $z$ and $x$ axes,
respectively. The last term accounts for precessional velocities.
Note that in a strict sense, we should replace
$\Omega_Ft$ by $\int\Omega_F dt$ since the precession frequency is
increased during the first 4 orbits of the simulations.
For simplicity, however, we keep this notation and understand that 
this replacement should be made.
The velocity and position vectors can be used to calculate 
orbital elements in the binary frame.
These are denoted  $a_i$, $e_i$, $\Omega_i$, $\alpha_i$, 
$\omega_i$, $f_i$ for the semi-major axis, eccentricity, longitude of ascending node,
inclination with respect to the binary plane, longitude of pericentre and 
true anomaly of particle $i$, respectively.
Since the planetesimals are initialised to be coplanar with the disc midplane,
and the equatorial plane of the spherical computation grid,
it follows that $\alpha_i=\gamma_F$ and $\Omega_i=\pi$ at $t=0$. 

Additionally,
we are interested in the inclination of the planetesimals with 
respect to the local disc midplane, which we will denote by the symbol $\delta$:
\begin{eqnarray}
\cos{(\delta_i)}&=&\frac{{\bf L}(r_i) \; . \; {\bf j}_i}{|{\bf L}(r_i)|\;|{\bf j}_i|} \nonumber\\
&=&\sin{(\alpha_i)} \sin{(\alpha_d)} \cos{(\Omega_i-\Omega_d)}+
\cos{(\alpha_d)} \cos{(\alpha_i)}
\label{relativeinc}
\end{eqnarray}
where ${\bf j}_i$ is the specific angular momentum vector of particle $i$, 
and ${\bf L}(r_i)$ is the angular momentum vector of the disc annulus
which has the same distance from the central star, $r_i$, as 
planetesimal $i$ at the current time. In this way, we
measure the inclination of the planetesimal with respect to the
local disc, which we will describe as the {\it relative
inclination} from now on.
The symbols $\alpha_d$ and $\Omega_d$ denote the inclination and nodal 
precession angles, respectively, of the local gas disc annulus with respect 
to the binary orbital plane.
Because the disc is rigidly precessing (and not strongly warped),
and because of our accurate choice of $\Omega_F$, the disc midplane
stays very close to the equatorial plane of the computational grid,
and we have approximately that $\alpha_d=\gamma_F$ and 
$\Omega_d=\pi+\Omega_Ft$ during the simulations.
If planetesimals stay close to the disc midplane 
(such that $|\Omega_i-\Omega_d|\ll 1$, $|\alpha_i-\alpha_d|\ll 1$),
Eq.~(\ref{relativeinc}) can be approximated by:
\begin{eqnarray}
\delta_i^2\simeq(\Omega_i-\Omega_d)^2 \sin^2{(\alpha_i)}+(\alpha_i-\alpha_d)^2.
\label{shortrelativeinc}
\end{eqnarray}
Note that such an approximation is generally valid for two orbits, 
whose orbital parameters ($\Omega_i$, $\alpha_i$) are very similar.

It will be instructive to understand which contributions 
(gas drag, disc gravity, binary companion) are
responsible for changing the orbital elements of the planetesimals. 
In order to measure this numerically, each force contribution is
transformed into the binary frame, where we take the sum of the direct and 
indirect parts
when considering the gravity of the disc or companion. We 
then calculate the radial $F_R$, tangential $F_T$ and normal
$F_N$ components with respect to the planetesimal orbit in 
the binary frame for each of the accelerations. These can be
used to calculate the changes of osculating orbital elements 
\citep{murray}. Here we will only state some of
them that will be used later for the discussion of results
(noting that a dotted quantity denotes its time derivative):
\begin{eqnarray}
\dot{\alpha_i}&=&F_N\cdot\frac{r_i}{j_i} \cos{(\omega_i+f_i)}\\
\dot{\Omega_i}&=&F_N\cdot\frac{r_i}{|{\bf j}_i-({\bf e}_3.{\bf j}_i){\bf e}_3|} 
\sin{(\omega_i+f_i)}.
\end{eqnarray}
For the change of relative inclination, 
differentiating Eq.~(\ref{relativeinc}) with respect to time gives:
\begin{eqnarray}
-\sin{(\delta_i)} \cdot \dot{\delta}_i &=& \cos{(\alpha_i)} \sin{(\alpha_d)}
\left[\dot{\alpha}_i \cos{(\Omega_i-\Omega_d)}-\dot{\alpha}_D\right]\nonumber\\
&+& \sin{(\alpha_i)} \cos{(\alpha_d)} 
\left[\dot{\alpha}_D \cos{(\Omega_i-\Omega_d)} -\dot{\alpha}_i\right]\nonumber\\
&+& \sin{(\alpha_i)} \sin{(\alpha_d)} \sin{(\Omega_i-\Omega_d)} 
(\dot{\Omega}_d-\dot{\Omega}_i).
\label{deltarate}
\end{eqnarray}
When calculating the change of relative inclination 
$\dot{\delta}_i$ we note that changes in the disc
inclination or nodal precession can only be caused by the binary 
companion. Hence when considering the change
due to drag force or disc gravity we set $\dot{\alpha}_d=0$ 
and $\dot{\Omega}_d=0$ in Eq.~(\ref{deltarate}).

\begin{table*}[t]
\begin{center}
\begin{tabular}{cl}
\hline
\hline
Variable & Variable description \\
\hline
$\gamma_F$ & Inclination between disc midplane and binary orbit plane\\
${\bf D}$  & Binary separation \\
$\omega_B$ & Binary orbit frequency \\
${\bf \Omega}_F$ & Precession frequency of the precessing frame and disc \\
$P_d$ & Orbital period at 20 AU \\
$M_d$ & The disc mass used in a simulation \\
${\hat M}$ & Our nominal disc mass (${\hat M}= 4.52 \times 10^{-3}$ M$_{\odot}$) \\
$h=H/R$ & Disc aspect ratio \\
$\alpha$ & Viscosity parameter \\
$\rho$ & Gas density \\
$s_i$ & Planetesimal physical radius \\
${\hat {\bf X}}$ & Vectors and coordinate quantities in the precessing frame 
denoted with the `hat' symbol \\
${\bf X}$ & Vectors and coordinate quantities in the fixed binary frame
denoted without the `hat' symbol \\
$a_i$ & Semi-major axis of planetesimal $i$ \\
$e_i$ & Eccentricity of planetesimal $i$ \\
$f_i$ & True anomaly of planetesimal $i$ \\
$\alpha_i$ & Inclination angle between binary orbit plane and orbit plane of
planetesimal $i$ \\
$\omega_i$ & Longitude of pericentre of planetesimal $i$ (also referred to as
the apsidal precession angle) \\
$\Omega_i$ & Longitude of ascending node of planetesimal $i$ (also referred to as
the nodal precession angle) \\
$\delta_i$ & Inclination of planetesimal $i$ relative to the local disc  \\
$\alpha_d$ & Inclination of local disc annulus relative to the binary orbit plane \\
$\Omega_d$ & Precession angle of local disc \\
$\alpha_i$ & Inclination angle between binary orbit plane and orbit plane of planetesimal $i$\\
\hline
\end{tabular}
\end{center}
\caption{Table of variable names.} 
\label{table1}
\end{table*}

\subsection{Collision detection}
In order to obtain collisional velocity distributions that are 
statistically relevant, it is necessary to accumulate data
on a large number of direct collisions between planetesimals.
Because we include the effect of disc gravity acting on the planetesimals,
which incurs a large computational overhead, we are only able to
integrate 50 planetesimals for each size that we consider. 
This means that we need to use an approximate method
for estimating the collision velocities between planetesimals, instead
of detecting direct collisions between them.

The approximation that we adopt in this work is to
treat the planetesimals not as individual particles, but
rather as representatives of their orbits, which we conceive of 
as being highly populated.
These orbits are assumed to have a circular cross section with a radius
$\Delta r=2\cdot10^{-4}$ AU in physical units
(note that this value has been used for the inflated radii of 
planetesimals in \cite{xie}, where a direct collision detection method
was used).
In order to estimate impact velocities for colliding planetesimals,
we detect the moments when two orbits defined by their
osculating elements intersect,
and estimate the velocity of impact that planetesimals at
the point of intersection would have. 
When running simulations which consider the general dynamics of planetesimals in
misaligned binary systems, we distribute the planetesimals
with a large range of semi-major axes within the protoplanetary disc
initially.
When running simulations which are designed to calculate the collisional velocities, 
however, we perform
separate runs in which the initial planetesimal orbits are
distributed with a narrow range of semi-major axes ($\Delta a=10^{-3}$ AU).
This is to maximise the number of orbit crossings, and hence improve
our collision statistics.

The condition for two orbits represented by
particles $i$ and $j$ to cross is given by the vector equation:
\begin{eqnarray}
{\bf r}_i(\phi_i,\Omega_i,\alpha_i,a_i,e_i,\omega_i)={\bf r}_j(\phi_j,\Omega_j,\alpha_j,a_j,e_j,\omega_j)
\label{vector_r_i}
\end{eqnarray}
where ${\bf r}_i$ is given by:
\begin{eqnarray}
{\bf r}_i=r_i(\phi_i,a_i,e_i,\omega_i) \left(\begin{array}{c} \cos{(\Omega_i)} \cos{(\phi_i)} -
\sin{(\Omega_i)} \sin{(\phi_i)} \cos{(\alpha_i)}\\ \sin{(\Omega_i)} \cos{(\phi_i)}
+ \cos{(\Omega_i)} \sin{(\phi_i)} \cos{(\alpha_i)} \\ \sin{(\phi_i)} \sin{(\alpha_i)}
\end{array} \right)\nonumber\\
\label{r3d}
\end{eqnarray}
with
\begin{eqnarray}
r_i(\phi_i,a_i,e_i,\omega_i)=\frac{a_i(1-e_i^2)}{1+e_i \cos{(\phi_i-\omega_i)}}=
\frac{a_i(1-e_i^2)}{1+e_i \cos{(f_i)}},
\label{r_i}
\end{eqnarray}
and similarly for the orbit represented by particle $j$.
These vector equations involve orbital elements which are
defined in the fixed binary frame.
The angles $\phi_i$ and $\phi_j$ in Eq.(\ref{vector_r_i}) are the
angular distances of the orbit crossing point to the nodal 
lines of orbits $i$ and $j$, respectively, where the nodal lines
are located in the binary orbit plane. For general crossing orbits,
where the semi-major axes and eccentricities of the two orbits
differ, it is not possible to solve Eq.~(\ref{vector_r_i}) directly.
For circular orbits with the same semi-major axis $\left(r_i=r_j\right)$,
however, we can solve Eq.~(\ref{vector_r_i}) for $\phi_i$ and $\phi_j$,
giving the values of these angles at the two points of
intersection for orbits $i$ and $j$:
($\phi_{j1}$, $\phi_{i1}$) and ($\phi_{j2}$, $\phi_{i2}$). These 
angle are defined by the expressions
\begin{eqnarray}
\tan{(\phi_{j1})}&=& \frac{\sin{(\Omega_i-\Omega_j)}}{\cos{(\alpha_j)} \cos{(\Omega_i-\Omega_j)}-
\sin{(\alpha_j)}/ \tan{(\alpha_i)}} \nonumber\\
\phi_{j2}&=&\phi_{j1}+\pi\nonumber\\
\cos{(\phi_{i1})} &=& \cos{(\phi_{j1})} \cos{(\Omega_i-\Omega_j)} \nonumber \\
 & + & \sin{(\phi_{j1})}
\cos{(\alpha_j)} \sin{(\Omega_i-\Omega_j)} \nonumber\\
\sin{(\phi_{i1})} &=&\frac{\sin{(\phi_{j1})} \sin{(\alpha_j)}}{ \sin{(\alpha_i)}}
\end{eqnarray}
The solution for $\phi_{i2}$ is obtained by using $\phi_{j2}$ 
instead of $\phi_{j1}$ in the last two equations.

The above solutions have been derived under the
assumption that the orbits are circular with the same semi-major axis. 
If they have a finite eccentricity and
different semi-major axes, we still assume that the point of closest 
approach is at these longitudes\footnote[1]{Note that this simplification was 
used to speed up the collision test in the simulations. 
Full numerical evaluation of the closest approach point of the two orbits gave 
good agreement with this assumption.}
$(\phi_{i1}$, $\phi_{j1})$ and  $(\phi_{i2}$, $\phi_{j2})$.
Hence we define orbital crossing or ${\it collision}$, if the
following condition is fulfilled:
\begin{eqnarray}
\left|r_i(\phi_{i1},a_i,e_i,\omega_i)-r_j(\phi_{j1},a_j,e_j,\omega_j)\right|\le\Delta r
\label{check}
\end{eqnarray}
where $r_i(\phi_{i1},a_i,e_i,\omega_i)$ is given by Eq.~(\ref{r_i}), 
and this condition is also checked for the other potential crossing point
($\phi_{i2}$, $\phi_{j2}$).\\
After every one hundred time steps
the condition (\ref{check}) is checked for the 11175 particle pairs 
that arise from the 50 particles integrated for each size.
Since this gives a total on the order of $10^7$ pairs for the simulated time
intervals we considered, we 
obtain quite a large number of pairs for which condition 
(\ref{check}) is fulfilled. Hence we are able to obtain
good statistics on collisional velocities despite the relatively low 
number of planetesimals modelled.

For orbit $i$ the velocity at the first orbit crossing location is:
\begin{eqnarray}
{\bf v}_{i1}&=&\frac{d}{dt}{\bf r}_i(\phi_{i1},\Omega_i,\alpha_i,a_i,e_i,\omega_i)=v_{ri}{\bf\breve{r}}_i+v_{\phi i}{\bf\breve{\phi}}_i\nonumber\\
&=&\sqrt{\frac{GM_\star}{a_i(1-e_i^2)}}\left[e_i \sin{(\phi_{i1}-\omega_{i})}
{\bf\breve{r}}_i+(1+e_i \cos{(\phi_{i1}-\omega_i)}){\bf\breve{\phi}}_i\right]\nonumber\\
\label{velocity}
\end{eqnarray}
and likewise for orbit $j$. Here ${\bf\breve{r}}_i$ is the particle 
unit vector calculated
from Eq.~(\ref{r3d}) and ${\bf\breve{\phi}}_i$ is the unit vector in
the $\phi$ direction. Likewise $v_{ri}$ and $v_{\phi i}$ are the 
velocity components in the $r$ and $\phi$ directions.
Note that this expression does not account for precessional velocities. However,
since they do not contribute to the relative velocities at orbital 
crossing points (where ${\bf r}_i={\bf r}_j$) they can be
safely omitted. The relative velocity at the first crossing location is 
then simply $\Delta v=|{\bf v}_{i1}-{\bf v}_{j1}|$,
and likewise for the second location ($\phi_{i2}$, $\phi_{j2}$). 
Later we will present averages of
collisional velocities calculated in this way.

\subsubsection{Validity of the orbit crossing technique}
An important question to address is under what conditions does the
orbit crossing technique that we have described above agree with
the direct detection of collisions when calculating average
collision velocities within a planetesimal swarm. 

The first point to note is that the method we use to
calculate the point of closest approach between two
orbits assumes that this occurs along the line
of intersection between the two orbit planes. For 
orbits which are mutually inclined with respect
to one another, this assumption is valid. But it
obviously breaks down for coplanar orbit planes,
and our method reports the wrong points of intersection
in this case.
So our method of detecting orbit crossing is
only strictly valid for mutually inclined orbits.

In order to address the more general question of whether the
orbit crossing method gives collisional velocities
which are similar to the direct collision detection method,
we have run a number of pure N-body simulations.
The first test we ran used a narrow ring of particles centred around 10 AU
with static orbital elements (no binary companion was included)
that were very similar to those reported in Fig. 7 of this paper at 60 orbits.
The direct detection simulation 
used 1000 particles and was run for approximately 2000 local orbits.
The orbit crossing simulation run used 50 particles
and was also run for approximately 2000 local orbits. We found that 
the mean collision velocities reported by each method agreed to
within approximately 30 \%. The second test was similar, but
included a binary companion inclined by $25^{\circ}$ to the
ring of planetesimals, which were initially placed on 
circular orbits. This was run for 2000 local orbits and
again good agreement was found between the two collision detection methods,
with the mean collision velocity reported by the orbit crossing
method being approximately 40 \% larger than that reported
by the direct detection method. We conclude that the orbit crossing
method gives fairly reliable results for the mean collision velocity in
a planetesimal swarm provided that the dominant contribution to
the relative velocities arises because of differential nodal precession.

\subsubsection{Analytical estimate of collisional velocities}
To gain a better understanding of the contributions that control the collisional 
velocities, we now derive an analytical estimate.
Consider two particle orbits A and B, which have very similar orbital elements
(i.e. $a_A=a_B$, $e_B=e_A+\Delta e$, $\Omega_B=\Omega_A+\Delta \Omega$,
$\alpha_B=\alpha_A+\Delta\alpha$, $\omega_B=\omega_A+\Delta\omega$), 
where the $\Delta$-quantities are assumed to be very small.
The above orbital elements are measured with respect to the binary orbit plane.
Consider now a coordinate system that is coplanar with the orbit of particle A.
The velocity of particle A is then given by Eq.~(\ref{velocity}) 
--  i.e. ${\bf v}_A=v_{rA}{\bf\breve{r}}_A+v_{\phi A}{\bf\breve{\phi}}_A$. 
With respect to the $x-y$ plane of this coordinate system, the 
orbit of particle B will be inclined by an angle $\delta_{AB}$, because its 
plane is slightly different due to the quantities $\Delta \alpha$ and
$\Delta\Omega$. But since these differences are very small by assumption, 
the relative inclination $\delta_{AB}$ between the two particle orbits is small. 
Note also that in such a coordinate system we can always choose 
$\Omega_A$ such that $\Omega_B=0$. Hence from Eq.~(\ref{r3d}) it follows that:
\begin{eqnarray}
{\bf\breve{r}}_B={\bf\breve{r}}_A+{\bf\breve{dr}}\hspace{0.5cm}{\rm with} \hspace{0.5cm}{\bf\breve{dr}}=
\left(\begin{array}{c}0\\0\\ \sin{(\phi_B)}\delta_{AB}\end{array}\right),
\label{rBrA}
\end{eqnarray}
where Eq.~(\ref{rBrA}) has been derived using the assumption that
the mutual inclination $\delta_{AB} \ll 1$.
Hence the orbit crossing point (${\bf r}_A={\bf r}_B$) 
corresponds to $\phi_B=0$, from which it follows that:
\begin{eqnarray}
{\bf\breve{dr}}={\bf
0}\hspace{0.5cm}and\hspace{0.5cm{\bf\breve{d\phi}}}=
\left(\begin{array}{c}0\\0\\\delta_{AB}\end{array}\right).
\end{eqnarray}
The velocity vector of orbit B at the orbit crossing point is thus:
\begin{eqnarray}
{\bf v}_B=(v_{rA}+dv_r){\bf\breve{r}}_A+(v_{\phi A}+dv_\phi)
({\bf\breve{\phi}}_A+{\bf\breve{d\phi}})
\end{eqnarray}
To first order we neglect the term $dv_\phi{\bf\breve{d\phi}}$, and 
the relative velocity between the two orbits at orbital crossing is:
\begin{eqnarray}
{\bf\Delta v}={\bf v}_B-{\bf v}_A=dv_r{\bf\breve{r}}_A+
dv_\phi{\bf\breve{\phi}}_A+v_{\phi A}{\bf\breve{d\phi}}.
\end{eqnarray}
Because
${\bf\breve{r}}_A.{\bf\breve{\phi}}_A=
{\bf\breve{r}}_A.{\bf\breve{d\phi}}={\bf\breve{\phi}}_A.{\bf\breve{d\phi}}=0$,
the relative velocity becomes:
\begin{eqnarray}
\Delta v^2=dv_r^2+dv_\phi^2+v_{\phi A}^2\delta_{AB}^2=
dv_r^2+dv_\phi^2+v_K^2\delta_{AB}^2.
\end{eqnarray}
In the second equality we can replace $v_{\phi A}$ by the keplerian velocity 
$v_K$ to this order.
From Eq.~(\ref{velocity}) the radial and azimuthal velocity 
differences are to first order:
\begin{eqnarray}
dv_r&=&v_r(e+\Delta e,\omega+\Delta\omega)-v_r(e,\omega)\nonumber\\
&\simeq&\sqrt{\frac{GM_\star}{a(1-e^2)}}
\left(\Delta e \sin{(\phi-\omega)} -e\Delta\omega \cos{(\phi-\omega)}\right)\nonumber\\
dv_\phi&=&v_\phi(e+\Delta e,\omega+\Delta\omega)-v_\phi(e,\omega)\nonumber\\
&\simeq&\sqrt{\frac{GM_\star}{a(1-e^2)}}
\left(\Delta e \cos{(\phi-\omega)}+e\Delta\omega \sin{(\phi-\omega)}\right).\nonumber\\
\end{eqnarray}
In the limit $e\ll 1$ the relative velocity therefore becomes:
\begin{eqnarray}
\Delta v^2=v_K^2\left(\Delta e^2+e^2\Delta\omega^2+\delta_{AB}^2\right).
\end{eqnarray}
The quantity $\delta_{AB}$ is defined within the coordinate system 
that is coplanar with orbit A.
Since $\Delta\Omega\ll 1$ and $\Delta\alpha\ll 1$ we can apply the 
approximation introduced in
Eq.~(\ref{shortrelativeinc}) to express $\delta_{AB}$ as:
\begin{eqnarray}
\delta_{AB}^2=\Delta\alpha^2+ \sin^2{(\alpha)}\Delta\Omega^2
\label{relative_alpha}
\end{eqnarray}
and the relative velocity becomes:
\begin{eqnarray}
\Delta v^2=v_K^2\left(\Delta e^2+e^2\Delta\omega^2+\Delta\alpha^2+\sin^2{(\alpha)}
\Delta\Omega^2\right).
\label{ana_vel}
\end{eqnarray}
This result shows that relative velocities are generated for a variety
of reasons. As in the coplanar case,
relative velocities can be generated due to different eccentricities $\Delta e$,
or phasing of pericentres $\Delta\omega$.
In three dimensions, relative velocities will also be caused
by different inclinations $\Delta\alpha$ or phasing of nodal lines 
$\Delta\Omega$. If $\Delta e=\Delta\alpha=0$, 
$\Delta\omega=\Delta\Omega\sim 1$ and
 $\sin{(\alpha)}\sim\alpha$ we recover the result 
from \cite{lissauer} for randomised orbits:
\begin{eqnarray}
\Delta v=v_K\sqrt{e^2+\alpha^2}
\label{lissauer}
\end{eqnarray}

\section{Effect of disc gravity on planetesimal dynamics: single star case}
In this section we present the results of simulations
of planetesimals that interact gravitationally
with the disc and central star only. The planetesimals
are on orbits that are inclined with respect to the disc midplane,
and we switch off the drag force.
The knowledge gained in this section will help to understand the results in 
later sections 
when the binary companion and the gas drag force are included in the model, and
allow us to isolate the effect that the disc gravity alone has on the dynamics
of planetesimals on inclined orbits.

\begin{figure*}
\begin{center}
\includegraphics[width=120mm,angle=0]{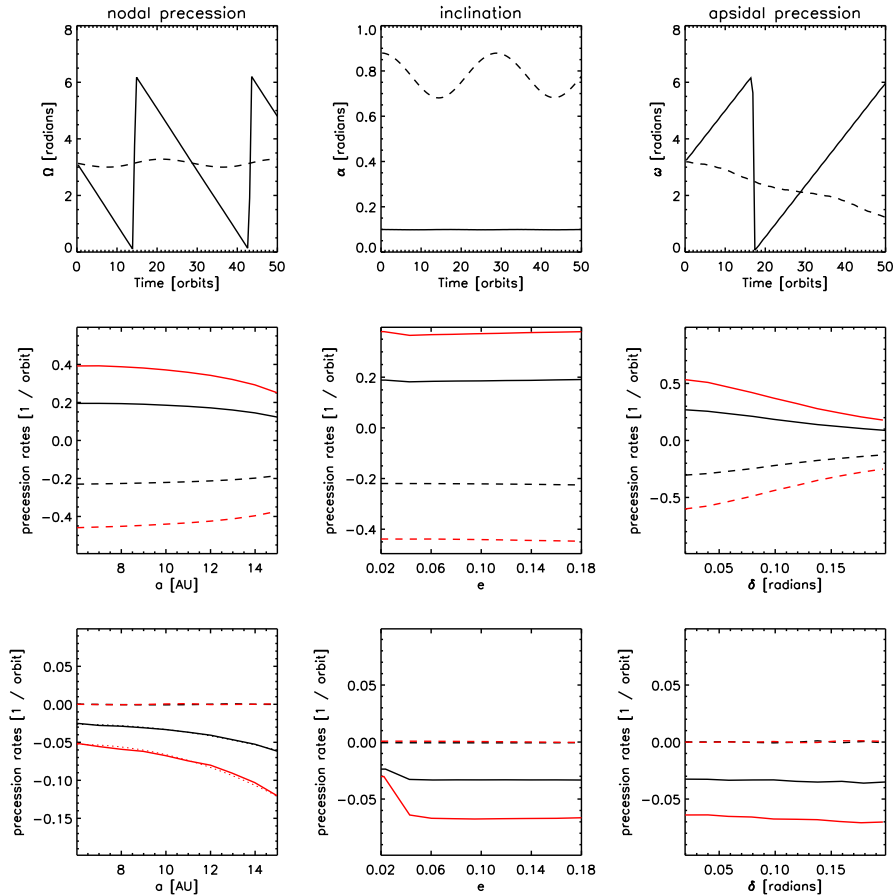}
\end{center}
\caption{Upper panels: nodal (panel 1) and apsidal precession (panel 3) angles as well 
as inclination angles
(panel 2) of a test particle experiencing the gravitational field of a disc and central
star only.
The solid lines show the results when viewed in a frame in which the disc is
assumed to be coplanar with the reference plane.
The dashed lines show the same quantities for the same model, but where the disc is now
treated as if it was inclined by $\gamma_F=0.78 \; (45^\circ)$ to the reference plane,
as it will be when the binary companion is included.
Middle panels: nodal (dashed lines) and apsidal (solid lines) 
precession rates when the disc is highly inclined
as a function of semi-major axis $a_i$ (panel 4), eccentricity $e_i$ (panel 5) 
and relative
inclination $\delta_i$ (panel 6) for a standard disc 
mass (black) and a disc with the double mass (red). Bottom
panels: nodal (dashed) and apsidal (solid) precession rates caused by the highly 
inclined disc as a function of
semi-major axis $a_i$ (panel 7), eccentricity $e_i$ (panel 8) and 
relative inclination $\delta_i$ (panel 9) for a
standard disc mass (black) and a disc with double that mass (red). 
In panel 7 a polynomial fit to the apsidal
precession rates is shown (dotted lines).} \label{od}
\end{figure*}

Throughout this paper we use the figure convention that the top left panel 
is referred to as panel 1, with
the remaining panels being labelled as 2, 3, 4... when moving from left 
to right and from top to bottom.
A simulation result for a reference particle is presented in 
Fig.\ref{od} (solid lines in panels 1-3 ).
This planetesimal has an initial semi-major axis of 10 AU, 
eccentricity of $e_i=0.1$ and relative inclination (inclination relative 
to the disc midplane) of
$\delta_i=0.1$. The quantities displayed are the nodal precession angle
$\Omega_i$ (panel 1), 
inclination $\alpha_i$ (panel 2) and apsidal
precession angle $\omega_i$ (panel 3). The quantities are calculated with 
respect to a reference plane that is
coplanar with the disc midplane ($\alpha_i=\delta_i$). We observe that the 
disc gravity causes retrograde nodal precession 
($\dot{\Omega_i}<0$) about the disc angular momentum vector,
and prograde apsidal precession ($\dot{\omega_i}>0$), 
while the inclination $\alpha_i$ remains unaffected.

We also studied planetesimals with different initial semi-major axes, 
eccentricities and relative inclinations.
The results are summarised in panels 4-6 of Fig.\ref{od}, which show the apsidal (solid)
and nodal (dashed) precession rates ($\dot \omega_i$ and $\dot \Omega_i$)
as a function of: semi-major axis $a_i$ (panel 4); 
eccentricity $e_i$ (panel 5);
relative inclination $\delta_i$ (panel 6). The black lines show results for a disc model
with the reference disc mass, and the red lines are for a model with twice the 
reference mass. 
As expected, the precession frequencies (nodal and apsidal) scale 
roughly with the mass of the disc.
Furthermore, the precession rates are larger in magnitude for particles 
that have smaller 
semi-major axes, as can be seen in panel 4.
Panel 5 shows that the precession rates (nodal and apsidal) are weakly
dependent on the eccentricity for the range in $e_i$ shown.
The dependence on relative inclination, however, is strong (see panel 6).

So far we have discussed the evolution of orbital elements measured with respect 
to a reference plane that is
coplanar with the disc midplane. Later on, however, we will consider 
planetesimal orbital elements 
measured with respect to the fixed binary plane,
which will be inclined with respect to the disc midplane. 
If the inclination of the
disc with respect to the binary plane is smaller than the relative inclination of
the particles ($\alpha_d\le\delta_i$), the nodal
and apsidal precession rates caused by the disc remain unchanged, and 
panels 4-6 of Fig.~\ref{od} apply.
If the inclination of the disc with respect to the binary plane 
becomes larger than the relative
inclination of the particles ($\alpha_d\ge\delta_i$), however, 
the nodal and apsidal precession rates will be different.

In order to understand the effect of the gravity of a disc 
that is highly inclined with respect to 
our reference plane,
we transform the planetesimal position and velocity vectors via the 
transformation given by Eq.~(\ref{trafo}), using a
representative inclination of the disc of 
$\alpha_d=\gamma_F=0.78 \; (45^\circ) > \delta_i =0.1 \; (5.7^\circ)$ 
and assume that the disc is
non-precessing with $\Omega_d=\Omega_F t=0$ (inclusion of a small precession
frequency $\Omega_F \ne 0$ does not change the result.)
The outcome of this transformation is shown by the dashed line in 
Fig.\ref{od} (panels 1-3) for the same
reference particle as described earlier. We observe that the nodal precession 
$\Omega_i$ angle (panel 1) and inclination angle
$\alpha_i$ (panel 2) are oscillating around a fixed value. 
This can be understood as follows. 
The angular momentum
vector of the particle still precesses around the angular momentum vector 
of the disc as before. 
Unlike before, however,
the disc has an inclination with respect to the reference plane of 
$\alpha_d=0.78 \; (45^\circ)>\delta_i$. 
Hence the precession
of the particle angular momentum vector causes the measured
inclination and nodal precession angles
to oscillate around those of the disc, with an amplitude given by the 
particle's relative inclination $\delta_i$.
We verify that for the reference particle the nodal precession rate measured with 
respect to the disc midplane is
$\dot{\Omega}\sim -0.23 \; (-13.2^\circ)$ per orbit, as seen from panel 4 
(black dashed line). 
This corresponds to a
precession period of 27 orbits, coinciding with the observed period 
of oscillation of the
inclination and nodal precession angles measured with respect to the reference plane 
(panels 1 and 2, dashed line).
Additionally, the particle appears to precess apsidally in a retrograde sense 
(panel 3, dashed line). 

The precession rates (nodal and apsidal) are displayed as functions 
of semi-major axis, eccentricity and relative
inclination in panels 7-9, respectively, for the highly inclined 
disc case, with the same line style and colour
convention as used before. As can be seen from these panels,
the net nodal precession rate is zero, 
since the disc
causes oscillation but no net change of the nodal precession angle.

From panels 8-9 we observe that the apsidal precession 
rate is roughly independent of eccentricity $e_i$
and relative inclination $\delta_i$. However it depends on the semi-major axis,
$a_i$, and also scales
roughly with the disc mass, as shown in panel 7.
For future purposes we fit this apsidal frequency by a second order polynomial:
\begin{eqnarray}
\dot{\omega}=\frac{M_d}{\hat{M}}\left[C_0+C_1
\left(\frac{a}{\rm AU}\right)+C_2\left(\frac{a}{\rm AU}\right)^2\right]\frac{1}{P_d}
\label{disc_aps}
\end{eqnarray}
with $C_0=-4.187\cdot 10^{-2}$, $C_1=5.118\cdot 10^{-3}$ and
$C_2=-4.242\cdot 10^{-4}$. $P_d$ is
the orbital period at the disc outer edge (which is nominally 
located at $r=10 \equiv 20$ AU),
$M_d$ is the disc mass and 
$\hat{M}$ is the nominal disc mass for the reference model introduced in Sect.~3.3.
The fit is displayed in panel 7 (dotted lines) and matches the numerical 
data (solid line) well.

To summarise, the measured effect of the gravity of an inclined disc on 
the planetesimal orbit depends on the ratio of the
inclination of the disc with respect to the binary plane, $\alpha_d$, and the 
inclination of the particle with respect to the
disc, $\delta_i$. If $\delta_i\ge\alpha_d$ the resulting nodal and apsidal 
precession rates caused by the disc are
displayed in Fig.\ref{od} (panels 4-6). If $\delta_i\le\alpha_d$ 
the precession around the disc angular momentum
vector causes oscillations in the nodal precession and inclination angles with an 
amplitude that is given by
$\delta_i$. The apsidal precession in this case is displayed in 
Fig.\ref{od} (panels 7-9) and can be approximated
by Eq.~(\ref{disc_aps}).

\section{Planetesimal dynamics in inclined binary systems}
\begin{table}[!h]
\begin{center}
\begin{tabular}{ccccc}
\hline
\hline
Model & $\gamma_F$ & $M_d$ & $D$ & $s_i$ \\
label & [degrees] & [$\hat{M}$] & [AU] & [m]\\
\hline
1&0&1&60&10000\\
2&25&1&60&10000\\
3&45&1&60&10000\\
&45&1&60&1000\\
&45&1&60&100\\
4&45&3&60&10000\\
5&45&6&60&10000\\
6&45&1&90&10000\\
7&45&1&120&10000\\
\hline
8&25&1&60&100\\
&25&1&60&1000\\
&25&1&60&10000\\
9&45&1&60&100\\
&45&1&60&1000\\
&45&1&60&10000\\
\hline
\end{tabular}
\end{center}
\caption{Table of runs: the first column gives the run label; the second column gives the 
inclination of the binary companion to the gas-plus-planetesimal disc, $\gamma_F$;
the third column gives the disc mass in units of the reference mass;
the fourth column gives the binary separation, $D$; 
the fifth column lists the planetesimal radii, $s_i$, included in the models.}
\label{table2}
\end{table}

We will now describe the dynamics of planetesimals that are orbiting 
in the full binary plus disc system.
The model parameters are summarised in table~\ref{table2}. The planetesimals are 
initially set-up on circular, keplerian 
orbits that are coplanar with respect to the disc midplane (i.e. $\delta_i=0$). 
They are distributed radially in the
interval  $a_i\in [6, 16]$ AU for models 1-7 (where
the disc truncation radius is $\simeq 20$ AU). In an additional 
series of runs we confine the planetesimals to a
narrow annulus with $\Delta a=10^{-3}$ AU centred around 
$a=10$ AU (models 8 and 9) to
maximize the number of collisions and study collisional velocities in more detail.

We treat model 3 as our reference case. The inclination of the binary companion to the
disc is initially set to $\gamma_F=0.78$ ($45^\circ$) and its separation from the central
object is set to $D=60$ AU. The disc mass used in the reference model is 
$M_d=1\hat{M}$ and corresponds to a
scaled minimum mass solar nebula as explained in Sect.~3.3. In the other models we
varied the initial inclination $\gamma_F$ of the binary companion to the gas-planetesimal disc 
(models 1-3), 
the disc mass $M_d$ (models 4 and 5) and the binary separation $D$ (models 6 and 7).

\subsection{Zero inclination case (model 1)}
\label{sec:zero-inc}
\begin{figure}[!t]
\begin{center}
\includegraphics[width=80mm,angle=0]{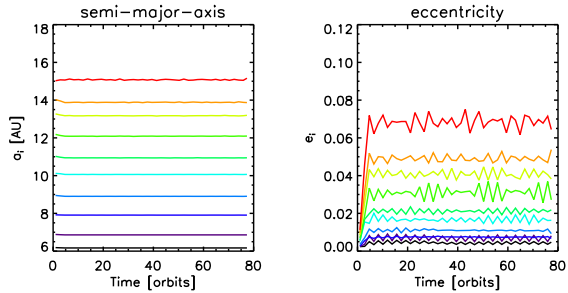}
\end{center}
\caption{Orbital elements for the coplanar case. Panel 1 shows the 
semi-major axes of different planetesimals,
where the colour convention introduced here is used in future plots. 
The eccentricity of the various planetesimals
is shown in panel 2. The binary separaion $D=60$ AU.}
\label{zeroinc}
\end{figure}
In this model the binary orbital plane is coplanar 
with the planetesimal-plus-disc midplane. 
From time $t=0$ the planetesimals experience the gravity of the disc
and gas drag forces (the planetesimals have size $s_i=10$ km),
and the mass of the binary companion is increased to its final value over a 
time of four orbits.
The evolution of the semi-major axes,
$a_i$, and eccentricities, $e_i$, are shown in Fig.\ref{zeroinc}. 
The colours correspond to different initial
semi-major axes, with darker blue colours representing the inner 
planetesimals and the green-red colours
representing planetesimals further out in the disc. We will also adopt this 
colour convention when
discussing simulation results later in the paper. 
As can be seen from Fig.\ref{zeroinc} (panel 1), 
the binary companion does not change the semi-major axes of the bodies,
as expected from secular theory.  
Eccentricities are generated, however, that are of the order of
($e_i\sim 10^{-2}$), as can be seen from Fig.\ref{zeroinc} (panel 2). 
Very modest eccentricities are generated because the planetesimals are orbiting
in a slightly non-keplerian potential due to the disc gravity. But we also see that the
eccentricities grow on a time scale of $\sim 5$ orbits due 
interaction with the binary whose mass grows over this time.
The values of eccentricity obtained are consistent with those
presented by \citet{Ciecielag}, who considered the evolution of
planetesimals in a gas disc with a circular binary companion,
but neglected the effects of disc gravity.
Because of the closer proximity to the
binary companion, the outer planetesimals are more strongly affected by these 
perturbations and their eccentricities are raised to larger values.
Note that the data in this and the following figures has been smoothed over a
temporal window of 1.8 orbits, but it is clear that after an
initial rise, the eccentricities remain essentially constant for
80 orbits (7155 years).

Secular perturbations from an eccentric, coplanar binary companion
lead to the generation of well-defined forced eccentricities,
and can also lead to strong
alignment of periastra for same-sized planetesimals
in the presence of gas drag
\citep{marzari,thebault}.
This causes a dramatic reduction
in the impact velocities for these planetesimals. But this effect is 
absent for a circular companion as considered here, and eccentricities are largely
generated by high frequency terms in the disturbing function. We find
that the periastra are not well aligned near the beginning of the
simulation, since a companion on a circular orbit cannot define a direction
of preferred alignment (this effect may be seen in Fig. 7 later in the paper
where we plot results for a simulation with a low inclination ($25^{\circ}$)
binary companion). This basic point is in agreement with results presented
by \citet{Ciecielag}, who also considered a circular binary. 
As such a binary companion on a circular orbit appears to
be singular in its effect on the orbits of planetesimals embedded in a gas disc.
This has potentially important consequences for the collisions between
planetesimals reported later in this paper.

\subsection{Low inclination case (model 2)}
\label{sec:low-inc}
\begin{figure}[!t]
\begin{center}
\includegraphics[width=80mm,angle=0]{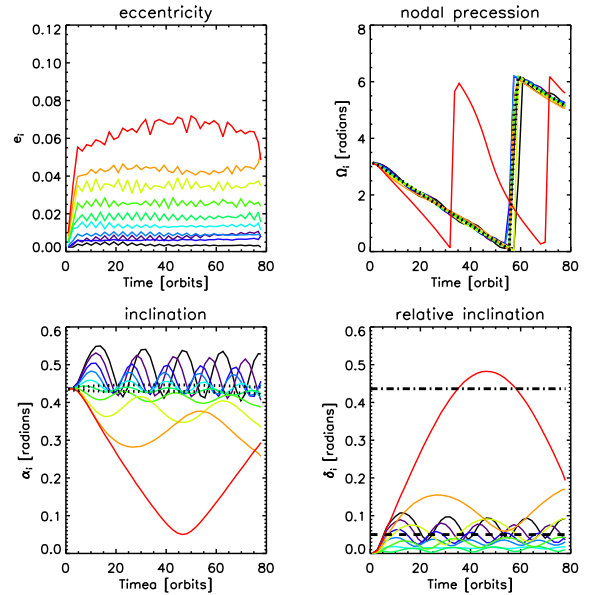}
\end{center}
\caption{Orbital elements for the low inclination
 case $\gamma_F=25^\circ$. The colours represent planetesimals at
different semi-major axes with the convention adopted from Fig.\ref{zeroinc}. 
The eccentricity is depicted in
panel 1. In panels 2 and 3 the nodal precession and inclination angles are shown, 
where the short dashed line
represents the inner and outer edge of the disc. In panel 4 the relative inclination with 
respect to the disc is shown,
where the short dashed line represents one pressure scale height and the 
dashed-dotted line indicates where
$\delta_i=\alpha_d$. The binary separation $D=60$ AU.} \label{lowinc}
\end{figure}

In this section we describe the simulation results for model 2,
for which only planetesimals of size 10 km were considered
and the binary inclination $\gamma_F=25^\circ$. 
Fig.\ref{lowinc} shows the evolution 
of eccentricities $e_i$ (panel 1),
nodal precession angles $\Omega_i$ (panel 2), inclinations $\alpha_i$ (panel 3) and relative
inclinations $\delta_i$ (panel 4) for the different planetesimals 
using the same convention of colours as 
described in Sect.~\ref{sec:zero-inc}.
As in the coplanar case, eccentricities are raised 
by interactions with the binary companion,
but because the binary orbit is circular the forced eccentricity predicted by
secular theory is negligible, and the eccentricities are generated by
high frequency perturbations.
We can observe that these are somewhat lower than in the coplanar case due 
to a reduced coplanar component of the companion's gravity.

\begin{figure}[h]
\begin{center}
\includegraphics[width=80mm,angle=0]{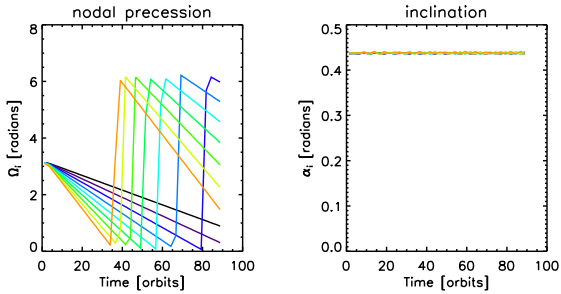}
\end{center}
\caption{Simulation results of planetesimals interacting with the binary system only
(no gas disc). 
The left panel shows the nodal precession, and the right panel shows the inclination.}
\label{nodalcomp}
\end{figure}

In panel 2 of Fig.\ref{lowinc} we see that most of the 
planetesimals precess nodally at a joint rate with the disc,
except for the outermost particle (red line), which will be discussed later
in this section. This joint precession can be 
explained by the presence of the disc, 
with its gravitational field playing the major role.
To illustrate this we also performed simulations of planetesimals 
that interact with the binary system only.
The corresponding nodal precession and inclination angles 
of the planetesimals for such a simulation are shown in Fig.\ref{nodalcomp}.
Due to the secular perturbation of the secondary star alone, 
the inclinations are expected to stay 
constant while the orbital planes of the planetesimals precess at their free 
particle rate \citep{papterquem}:
\begin{eqnarray}
{\dot \Omega}_i^{free}=-\frac{3}{8\Omega_K}\frac{GM_B}{D^3}(3\cos^2{\alpha_i}-1).
\label{free}
\end{eqnarray}
which is a strong function of semi-major axis, as $\Omega_K\sim r^{-\frac{3}{2}}$ 
is the keplerian angular velocity.
The left panel of Fig.\ref{nodalcomp} shows that the nodal 
lines of the planetesimals become progressively randomised, 
leading to the dispersion of the 
planetesimal disk into a cloud of bodies surrounding the star, 
as found by \cite{marzari2}.
Comparing this behaviour with the results displayed in Fig.\ref{lowinc} (panels 2 and 3) 
we can see that the presence of the disc and its gravitational field
causes fundamentally different behaviour of the planetesimals.
The planetesimals remain coupled to the disc, as their nodal 
precession angles stay close to 
the disc precession angle (dashed line in panel 2 of Fig.~\ref{lowinc}).
The role of gas drag in determining the relative inclination
between the disc midplane and the planetesimals is negligible for
this run for which the planetesimal size is 10 km. But for
smaller planetesimals gas drag does play a role, as we will describe
later in the paper.

\begin{figure}
\begin{center}
\includegraphics[width=80mm,angle=0]{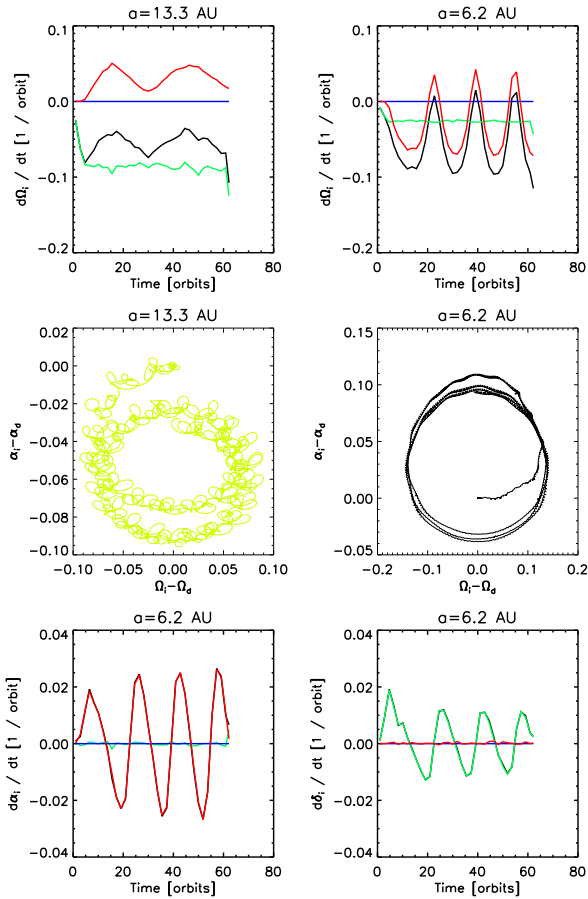}
\end{center}
\caption{Panels 1 and 2 show the nodal precession rates of a planetesimal with semi-major 
axis at $a_i=13.3$ AU (panel 1) and 
$a_i=6.2$ AU (panel 2). Shown are the contributions of the disc 
gravity (red line), binary companion (green
line), gas drag (blue line) and total rate (black line). 
In panels 5 and 6 we also show the rates of inclination change
with respect to the binary and disc midplane respectively for the inner particle. 
In panels 3 and 4 the trajectory of the tip of the
particle's angular momentum vector is shown projected onto
the ($\Omega_i-\Omega_d$, $\alpha_i-\alpha_d$) plane for the
outer (panel 3) and inner particle (panel 4).} \label{forces}
\end{figure}

In Fig.\ref{forces} we show the time
evolution of the nodal precession rates for the particle at 
$r=13.3$ AU (panel 1) and $r=6.2$ AU (panel 2). The
various curves represent contributions due to gas drag (blue line), 
disc gravity (red line), 
binary companion (green line) and
the total rate (black line) as calculated in Sect. 3.4. 
Because of its different radial location, the outer
particle orbit precesses faster in the retrograde sense than the inner particle 
if the disc is absent. 
Confirming this we can observe in
Fig.\ref{forces} (panels 1 and 2) that $\dot{\Omega}_i$ 
due to the companion (green line) is about
$\dot{\Omega}_i=-0.09 \; (-5.15^\circ)$ per orbit for the outer body,
and $\dot{\Omega}_i=-0.03 \; (-1.7^\circ )$ per orbit for the inner
body. It is the disc gravity, however, that compensates for this effect and causes the 
planetesimals to precess at a common net rate.
Since the disc precesses rigidly at a rate 
$\dot{\Omega}_D\simeq\Omega_F=-0.06 \; (-3.4^\circ )$ per orbit, which
corresponds to the free particle rate at about $a_i=11.5$ AU, disc gravity (red line) 
gives a positive contribution to the
precession rate for the outer particle ($\dot{\Omega}_i>0$) 
and a net negative contribution for the inner
particle ($\dot{\Omega}_i<0$). Since the particle-plus-disc 
system precesses in the retrograde sense the outer
particle precession is slowed down, while the inner particle 
precession is speeded up by the disc, 
such that both bodies precess together with the disc on average
at the rate $\Omega_F$. This result illustrates
the fundamental importance of including the effects of disc gravity when calculating
the dynamics of planetsimals in misaligned binary systems.

The nodal precession and inclination angles of the planetesimals 
show oscillations, as seen in
Fig.\ref{lowinc} (panels 2 and 3). 
These are caused by the disc gravity,
as explained in Sect. 4 and illustrated by the 
red line in panel 5 of Fig.~\ref{forces}. The
planetesimals effectively precess around the disc angular momentum vector
at the same time as the disc-plus-planetesimal system precesses
around the binary angular momentum vector.
Initially the planetesimals are coplanar with the disc midplane ($\delta_i=0$),
but as the system evolves during early times, differential nodal 
precession between particle $i$
and the disc occurs, leading to a build up of relative inclination $\delta_i$, since
$\delta_i^2\propto(\Omega_i-\Omega_d)^2$ as shown by Eq.~(\ref{shortrelativeinc}). 
The differential precession is greatest for planetesimals whose semi-major axes deviate 
most from $a=11.5$ AU (the radius at which planetesimals naturally want to precess 
at the same rate as the disc), so the relative inclination is largest for 
particles that are furthest
inside (black line) or outside (yellow line) this value, 
as demonstrated in panel 4 of Fig.~\ref{lowinc}. 

The disc model has an aspect rato $H/R=0.05$, so that planetsimals
which develop relative inclinations $\delta_i > 0.05$ will spend large fractions
of their orbits away from the disc midplane. Planetesimals interior to $8$ AU and 
exterior to $13$ AU have relative inclinations $\delta_i \simeq 0.1$, and so
spend large fractions of their orbits essentially
outside of the disc.
We also observe that the oscillation periods of the inclination, 
$\alpha_i$, and nodal precession angles,
$\Omega_i$, varies among the planetesimals. 
This is expected due to their different semi-major axes
and relative inclinations, and the observed periods are all consistent 
with Fig.\ref{od} 
(dashed lines in panels 4-6), which we recall shows how the disc alone acts on the
planetesimals.

We note that the inclination oscillations, seen in Fig.\ref{lowinc} (panel 3), 
begin in opposite senses for planetesimals whose 
semi-major axes are above or below $a=11.5$ AU 
(i.e. planetesimals below $a=11.5$ AU are perturbed onto
higher inclination orbits, while planetesimals exterior to $a=11.5$ AU are 
perturbed onto lower inclination orbits).
Panel 3 of Fig.\ref{forces} plots the inclination angle difference 
$\alpha_i-\alpha_d$ versus
the precession angle difference $\Omega_i-\Omega_d$ for a planetesimal
located at 13.3 AU. A similar plot for a planetesimal at 6.2 AU is shown
in panel 4. Each plot shows the trajectory of the tip of the planetesimal 
angular momentum vector relative to the disc angular momentum vector
(which is located at the origin).
For the inner particle the companion 
induces a differential precession
$\Omega_i-\Omega_d>0$ and the anti-clockwise precession around the 
disc angular momentum vector causes the
planetesimal to approach a higher inclination orbit 
$\alpha_i-\alpha_d>0$ during the first half of this
precession cycle. For the outer planetesimal the induced differential precession is 
$\Omega_i-\Omega_d<0$, and the anti-clockwise
precession causes perturbation onto a lower inclination orbit $\alpha_i-\alpha_d<0$.

We now return to the planetesimal that is orbiting at 
$a=15$ AU. This body shows quite distinct behaviour 
from all the other bodies, as shown in
Fig.\ref{lowinc} (red lines). This particle is 
dominated by the gravity of the companion, and
it nodally
precesses fast enough to decouple from the disc. 
As it precesses away from the disc, the 
relative inclination $\delta_i$ grows (panel 4 of Fig.\ref{lowinc}).
The precession rate
around the disc angular momentum vector is a decreasing function 
of relative inclination, so the precession period
becomes long for this particle. Because
$\Omega_i-\Omega_d<0$, we observe the
quasi-monotonic decrease of inclination in panel 3 of Fig.\ref{lowinc} 
until the reversal point at 
$t=47$ orbits, when
the differential nodal precession is $\Omega_i-\Omega_d=\pi$. 
The planetesimal orbital plane then
approaches the disc midplane from the other side ($\Omega_i-\Omega_d>0$),
and we observe the
quasi-monotonic increase of inclination after $t=47$ orbits. 

Over longer time scales than we have been able to consider
because of the computational expense of running the simulations,
we expect the outermost planetesimals which decouple from the
disc to show continued oscillations in their inclinations, with an oscillation period 
equal to the synodic precession period. Ignoring the possible
growth of planetesimals into planets, or their collisional
destruction, those bodies which remain 
almost coplanar with the disc should continue to do so over long
time scales until
the disc mass decreases due to viscous evolution 
and/or photoevaporation. 
Significant reduction of the disc mass would eventually allow
the planetesimal dynamics to become dominated by the companion star,
and they would decouple from the disc.
But, we also note that that the disc and binary orbit
plane will evolve toward alignment on the viscous evolution time 
at the outer edge of the of the disc
\citep{Terquem, larwood, fragner}. So it is possible that an initially
misaligned protoplanetary disc may align significantly over its lifetime,
bringing with it any planetary system which has formed within it.
The final degree of misalignment for a planetary system formed in an
inclined protoplanetary disc may therefore be substantially less than
the initial misalignment of the protostellar disc.

\subsubsection{Collisional velocities in low inclination case (model 7)}
\label{sec:low-inc-coll}
In this section we present the results of a simulation which examines
collisions between planetesimals in a system where the binary
companion has an inclination of $\gamma_F=25^\circ$ relative to
the disc midplane. To increase the collision rate to statistically meaningful
values which can be measured in a simulation, 
we initialise the planetesimals with a narrow range of semi-major axes ($\Delta
a=10^{-3}$ AU) centred around $a=10$ AU. We consider three different planetesimal 
sizes (100m, 1km and 10km),
and for each size we include 50 particles. We check the condition for orbital
crossing given by Eq.~(\ref{check}) every 100 time steps for each of the 11175
particles pairs (100 time steps corresponds to 0.011 orbital periods 
at 10 AU, the orbital radius of the planetesimals). If the 
orbit crossing condition is satisfied, then we 
calculate the velocity at the crossing point for both planetesimal
orbits according to Eq.~(\ref{velocity}). 

\begin{figure*}[t]
\begin{center}
\includegraphics[width=160mm,angle=0]{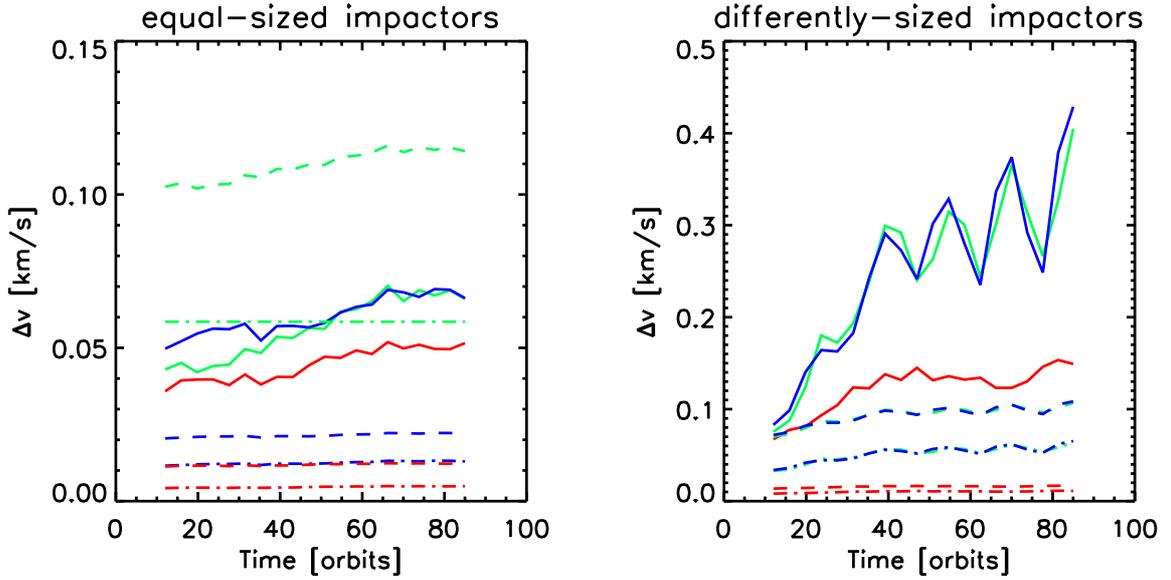}
\end{center}
\caption{Average collisional velocities in the low inclination 
$\gamma_F=25^\circ$ case between equally sized
(left panel) and differently sized planetesimals (right panel) for
planetesimals centred around 10 AU from the central star. 
Left panel: collisions between 10km-10km bodies (green-solid); 1km-1km
(blue-solid); 100m-100m (red-solid). Right panel: collisions between 10km-1km bodies
(green-solid); 10km-100m (blue-solid); 1km-100m
(red-solid). Threshold velocities for catastrophic disruption corresponding to 
the different size-combinations
for the weak (dash-dotted line) and strong aggregates (dashed line) are also shown.}
\label{vel_lowinc}
\end{figure*}

Before discussing the results of model 7, it is worth recapping
what we might expect based on previous work in which the gravity
of the disc was neglected. Same-sized planetesimals being
perturbed by an eccentric, coplanar binary companion will experience
a growth in their forced eccentricity, but gas drag damping will cause
strong orbital phasing dramatically reducing collisional velocities
\citep{marzari, thebault}. The different phasing of pericentres for
planetesimals of different size, however, leads to large collision velocities
which are likely to be disruptive. The inclusion of a small inclination
($\alpha_i \le 5^{\circ}$) causes different sized planetesimals to orbit
in different planes, such that collisions between similar sized
bodies are more frequent than between different sizes. The fact that
the pericentre phasing is maintained in this scenario means that 
planetesimal growth may be more likely to occur in inclined systems
\citep{xie}.

The collision velocities we obtain are shown in Fig.\ref{vel_lowinc} (solid lines) for 
collisions between equally sized (panel 1) and differently sized planetesimals (panel 2).
As mentioned in Sect.~\ref{sec:zero-inc}, an important difference
between our set-up and previous work is that we consider a circular,
inclined binary companion, resulting in a broad distribution of 
planetesimal longitudes of pericentre, $\omega_i$, even for planetesimals
of the same size. Consequently, we see that the collision velocities
for equal sized bodies in panel 1 are quite large, being 
between 50 - 70 ms$^{-1}$ at the end of the simulation. 
Although it appears that the circular binary is largely
responsible for the growth of eccentricity and the misaligned
periastra of orbits for same sized planetesimals, it is possible
that gravitational perturbations associated with spiral waves in the
disc also contribute.
The collision
velocities for differently sized bodies, however, are even larger than
for same-sized bodies,
and exceed 400 ms$^{-1}$ after 80 orbits (see panel 2).

\begin{figure}[h]
\begin{center}
\includegraphics[width=80mm,angle=0]{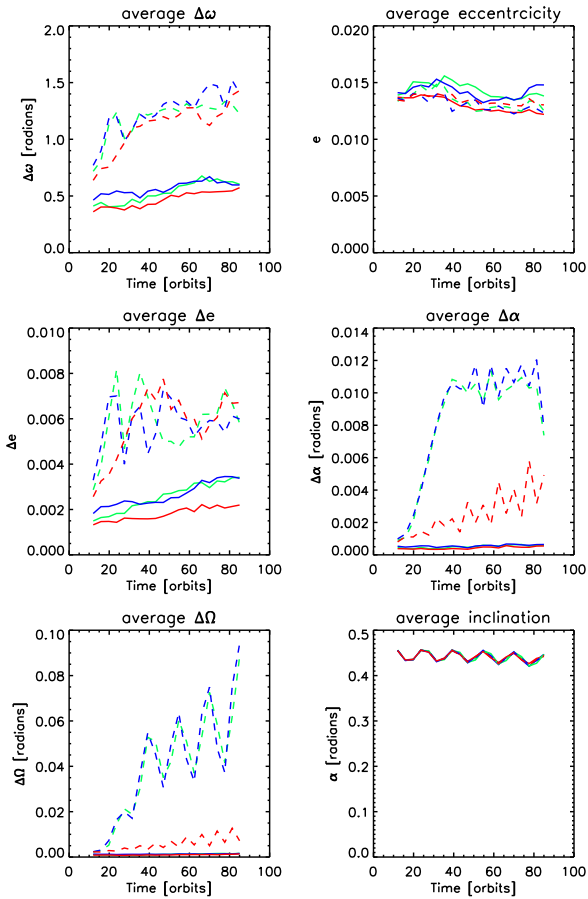}
\end{center}
\caption{Average orbital parameters of the colliding planetesimal pairs in the
low inclination
$\gamma_F=25^\circ$ case. The line colours and styles are as follows: green-solid
(10km-10km); blue-solid
(1km-1km); red-solid (100m-100m); green-dashed (10km-1km);
blue-dashed (10km-100m); red-dashed (1km-100m).}
\label{orbital_lowinc}
\end{figure}

Fig.\ref{orbital_lowinc} shows averages of all the quantities 
that determine the relative velocities
according to the analytical estimate given by Eq.~(\ref{ana_vel}),
and we can use this data to determine the main contributors to
the collision velocities shown in Fig.\ref{vel_lowinc}.
Panel 1 shows the difference of longitude of pericentre
$\Delta\omega_i$, panel 2 shows the average eccentricity $e$,
panel 3 shows the difference in eccentricity $\Delta e_i$,
panel 4 shows the difference in inclination $\Delta\alpha_i$,
panel 5 shows the angle between the nodal lines $\Delta\Omega_i$ and
panel 6 shows the average
inclination $\alpha_i$. Note that these averages were obtained only
for orbits which mutually cross, and do not represent the distributions
for the whole ensemble of particles.
Using the numbers extracted from these figures in 
Eq.~(\ref{ana_vel}) we can reproduce the relative velocities shown 
Fig.\ref{vel_lowinc} with good accuracy,
indicating that Eq.~(\ref{ana_vel}) is a valid approximation
and that our collision detection technique is generating
collision velocities which are consistent with expectations.

For collisions between planetesimals of the same size the dominating contribution 
to the square of the relative velocity comes from the term 
which is proportional to $e_i^2\Delta\omega_i^2$, 
while all other terms are at least an order of magnitude
smaller. This implies that planetesimals with the same size all orbit more or 
less in the same plane (as shown by panels 4 and 6), and relative velocities
are generated due to misalignment of their orbit pericentres, $\Delta\omega_i$,
a result which is broadly consistent with those obtained by \citet{xie}.
We note that our collision detection method is not strictly valid in
this regime, because it predicts the locations of the orbit crossing
points incorrectly for coplanar orbits. But, because the average 
pericentre misalignment is on the order of unity (measured in radians),
the average collision velocities reported by the algorithm are consistent with the
analytical prediction $v_{coll} \simeq e \, \Delta \omega$ to within a factor
of order unity. Under these conditions, incorrect detection of the orbit crossing
locations still leads to estimates of the {\it average} collision velocities being
approximately correct.

Although panel 1 of Fig.\ref{orbital_lowinc} shows that the misalignment
of pericentres increases slightly over time when considering only
those particles whose orbits intersect, we find that 
the average pericentre misalignment calculated over all same-sized particle
pairs actually decreases during the simulation. The reason for this
is not clear, but it is interesting to note that even when the binary orbit
is circular, the long term evolution appears to be toward one where
the pericentres tend toward alignment, and this effect is most marked
for the smallest planetesimals suggesting that it is an effect due to
gas drag.

As seen in panel 2 of Fig.\ref{vel_lowinc}, planetesimals of 
different sizes tend to have much larger collisional velocities.
One reason is the increased misalignment of pericentres, 
as can be observed in
Fig.\ref{orbital_lowinc} (panel 1, dashed lines). 
This may be a direct consequence of gas drag, as 
discussed in previous work \citep{marzari, thebault},
but it may also be due to the fact that particles
which experience stronger gas drag orbit closer to
the disc midplane than larger planetesimals do. This
then affects the gravitational perturbations experienced
by the particles as they travel through the disc, which
may increase the pericentre misalignment. But a more important
contribution to the
square of the relative velocities comes from the term $\propto \sin^2{(\alpha_i)}
\Delta\Omega_i^2$. We find that due to the different gas drag strengths,
planetesimals of different sizes tend to precess nodally at different rates for some 
initial time until disc gravity causes them to precess together with the disc.
While smaller
planetesimals tend to precess together with the disc immediately and consequently 
their relative inclinations stay low, larger planetesimals tend to precess at their own free
particle rates for a longer time, causing a larger build up of relative inclination.
In other words, although the planetesimal swarm as a whole is forced on average
to precess with the disc by the disc gravity, smaller planetesimals oscillate
about the midplane of the disc with a smaller amplitude than larger planetesimals.
This induces a large misalignment of their orbital planes $\Delta\Omega_i$ as seen in 
panel 5 of Fig.\ref{orbital_lowinc} (dashed lines)
in addition to the misalignment of their pericentres. 

The frequency of collision between bodies is an important factor
during the growth of planetesimals, and \citet{xie} have suggested that
the differential phasing of nodal lines for planetesimals of different
sizes may increase the relative importance of collisions between
similar sized objects rather than different sized bodies. We have examined the
collision frequency reported by our collision detection technique by
simply counting the number of orbit crossing events reported during the
simulation. In basic qualitative agreement with \citet{xie}, we
find that the collision frequency during the simulation between
same-sized objects is a factor of 3 - 6  times larger than between
different sized objects. Our results thus support the idea that planetesimal
growth in inclined binaries is likely to proceed via accretion of similar
sized bodies.

We conclude that relative velocities between differently sized planetesimals 
tends to be large in binary systems
whose orbital plane is misaligned with the planetesimal-plus-disc plane, 
while relative velocities between
same sized planetesimals are largely unaffected by this non-coplanarity. 
But the circular orbit of the binary companion prevents strong
pericentre alignment for similar sized bodies, leading to significant
collisional velocities in this case too, in contrast to the situation
observed when the companion is on a coplanar eccentric orbit.
The question of what happens in the case of an eccentric, inclined
companion unfortunately goes beyond the scope of this paper but should
obviously be the focus of future work.

We now want to determine whether the collisional velocities seen in 
Fig.\ref{vel_lowinc} will lead to growth
or erosion of the planetesimals. 
In principle there are three possible outcomes: accretion,
in which the largest body involved in the collision contains more mass
after the collision; catastrophic disruption, in which more than half
of the total mass of the system is dispersed after the collision;
erosion (or cratering), where the largest body loses a small amount of
mass during the collision. Only accretion leads to the growth of a 
planetesimal.

For simplicity in interpreting our results, we adopt the universal law 
for the largest remnant mass from \cite{stewart}:
\begin{eqnarray}
\frac{M_{lr}}{M_{tot}}=1-\frac{1}{2}\frac{Q_R}{Q_{D}},
\end{eqnarray}
where $M_{lr}$ is the mass of the largest post-collision remnant,
and $M_{tot}=M_1+M_2$ is the total mass of the
two colliding bodies $M_1$ and $M_2$. The quantity 
$Q_R=0.5\frac{M_1M_2}{M_{tot}^2}\Delta v^2$ is the reduced mass
kinetic energy scaled by the total mass of the colliding system. 
For equally sized bodies accretional collisions
require $\frac{M_{lr}}{M_{tot}}>\frac{1}{2}$, from which it follows that 
$Q_R<Q_{D}$. Hence $Q_{D}$ is called the
catastrophic disruption limit of the collisions 
(the energy required to disperse half the total mass).
If $Q_R>Q_{D}$ collisions between equally sized bodies lead to catastrophic
disruption.
When considering collisions between differently sized bodies
the condition has to be modified. Let
$M_1\gg M_2$ such that $M_2=\mu M_1$ with $\mu\ll 1$. 
The condition for accretion is now $M_{lr}>M_1$, from which it
follows that $Q_R<\left(\frac{2\mu}{1+\mu}\right)Q_{D}$, which 
implies for the relative velocity threshold:
\begin{eqnarray}
\delta v_{D}=2\sqrt{1+\mu}\sqrt{Q_{D}} \label{limitvelocity}
\end{eqnarray}
The disruption limit $Q_{D}$ is sensitive to factors that influence the 
energy and momentum coupling between the
colliding bodies (i.e. impact velocity, mass ratio and material 
properties such as strength and porosity).
\cite{stewart} show that the catastrophic disruption curve 
$Q_{D}$ can be fit to an analytical
formula \citep{houssen}:
\begin{eqnarray}
Q_{D}=\left[q_SR_{12}^{9\mu_M/(3-2\phi_M)}+q_GR_{12}^{3\mu_M}\right]V_I^{(2-3\mu_M)} \label{limitQ}
\end{eqnarray}
where $\mu_M$ and $\phi_M$ are material properties. In this expression 
$R_{12}$ is the spherical radius of the combined
mass $M_{tot}$ assuming a density of $\rho_s=1$ g cm$^{-3}$. 
Since we use a density of $\rho_s=2$ g cm$^{-3}$ this radius is given
by $R_{12}=2^\frac{1}{3}(1+\mu)^\frac{1}{3}R_1$, 
where $R_1$ is the spherical radius of the larger mass $M_1$
involved in the collision. The first term on the right hand side of 
Eq.~(\ref{limitQ}) represents the strength
regime, while the second term represents the gravity regime. 
Very small bodies are supported by material
strength (first term), which decreases as the planetesimal size increases. 
As gravity becomes more important for
larger bodies the second term becomes dominant and increases
the disruption limit again.
\begin{figure}[t]
\begin{center}
\includegraphics[width=80mm,angle=0]{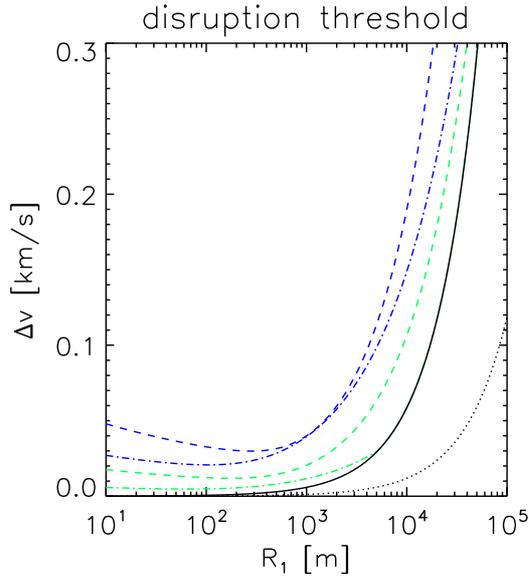}
\end{center}
\caption{Threshold velocities for collisions between equally sized planetesimals as 
a function of their
spherical radius $R_1$. The disruption curves for the weak 
(dashed-dotted line) and strong rock (dashed line)
are shown for two different impact velocities $V_I=50$m/s (green) and 
$V_I=2$km/s (blue) along with the solution
for rubble piles (black-solid line). The bodies escape velocity 
(black-dotted line) is also shown. }
\label{threshold}
\end{figure}
\cite{stewart} derive the constants $q_S$, $q_G$, $\mu_M$ and $\phi_M$ 
for weak aggregates, such as weak rock
and porous glass, by fitting the disruption curve to their numerical results. 
For weak aggregates they find (in cgs units)
$q_S=500$, $q_G=10^{-4}$, $\mu_M=0.4$ and $\phi_M=7$. 
For strong rocks they use the basalt laboratory
data and modelling results from \citet{benz}, and find 
$q_S=7\cdot 10^4$, $q_G=10^{-4}$, $\mu_M=0.5$ and
$\phi_M=8$. In the limit of very large bodies the disruption curve 
can be best fitted by their
results of colliding rubble piles. In this regime \cite{stewart} find
\begin{eqnarray}
Q_{RP}=1.7\cdot 10^{-6}R_{12}^2
\end{eqnarray}
for equal mass bodies and a factor of about three larger 
than this if the impactor size is much smaller than the target size.
For illustrative purposes we present the relative velocity threshold 
$\delta v_{D}$ obtained by combining
Eq.~(\ref{limitvelocity}) and Eq.~(\ref{limitQ}) in the case of identical 
colliding bodies ($\mu =1$) in Fig.\ref{threshold}.
Note that in the limit of large bodies at low impact velocities, 
we set $Q_{D}=Q_{RP}$, because the rubble-pile
solution for equal mass bodies defines a lower limit for disruption in this regime.
The figure shows the limiting velocity below which collisions will lead to net 
accretion as a
function of the planetesimal radius $R_1$ for the weak
(dashed-dotted line) and strong aggregates (dashed line) at two different 
impact velocities $V_I=2$km/s (blue line) and
$V_I=50$m/s (green line). It is apparent that the threshold velocity 
is increased for larger impact velocities as
more energy is partitioned into shock deformation. In contrast, 
at low impact velocities the momentum coupling
between the colliding bodies is more efficient and less energy is needed 
to cause erosion of planetesimals. As
gravity becomes more important than strength for very large bodies 
the curves join onto the disruption curve for
colliding rubble piles (black solid line). In this regime the threshold
velocity is about a factor of 3-5 larger than the escape velocity from the 
bodies surface (dotted line). \\
To compare the relative velocities obtained from our numerical simulations with 
the disruption threshold
velocity, we evaluate Eq.~(\ref{limitvelocity}) for equally ($\mu =1$) and 
differently sized $\mu =10^{-3}$ planetesimals
using $V_I=\Delta v$ in Eq.~(\ref{limitQ}) and choosing $R_1$ as the bigger of the 
two planetesimals involved in the collision.
If the rubble-pile disruption limit gives a higher estimate for 
a given size combination we use this limit instead.

We note that \cite{stewart} considered target-projectile mass
ratios only down to 0.03, instead of the values $10^3$ and $10^6$ that
apply to collisions of 1km or 100m sized bodies with a 10km sized
target. As such we are applying the \cite{stewart} results in a regime
where catastrophic disruption is unlikely to occur, and therefore 
outside of the regime of validity of their study, strictly speaking.
Nonetheless, it reasonable to assume that even if catastrophic disruption
does not occur during high velocity impacts, accretional growth is also
unlikely due to erosion.
 
The results are plotted in Fig.\ref{vel_lowinc} for the strong 
(dashed line) and weak aggregates
(dashed-dotted line). Except for collisions between equally 
sized $10$km bodies (green), the collisional
velocities are always substantially larger than the erosion/disruption
threshold. Hence we conclude that growth of
planetesimals probably can not occur for planetesimal sizes $<10$km. 
For the equally ($10$km) sized bodies on the other
hand the situation is not as clear. Although the collisional velocities lie below 
the strong aggregate threshold
(green dashed line) they are also slightly above the weak aggregate threshold 
(dashed-dotted line). This suggests
that collisions might also lead to erosion in this size regime. 
But it should be noted that we have not included a realistic 
planetesimal size distribution in these simulations, and it is possible
that collisions between planetesimals clustered in size around 10km,
whose orbit planes are very modestly inclined with respect to
one another, may allow accretion to occur for strong aggregates.
Examination of this possibility will require future simulations
that adopt a more realistic and continuous size distribution.

If collisions between $10$km sized
planetesimals do lead to growth then this will most definitely not occur 
in the standard runaway regime, because the
escape velocity is $\sim 10$m/s. It is possible, however, that 
type II runaway growth may occur \citep{kortenkamp} in which
the large velocity dispersion causes growth to be orderly while
the bodies remain small, but enters a runaway phase when large bodies
form whose escape velocities are larger that the velocity dispersion.
We conclude
that in an inclined binary system on a circular orbit,  
relative velocities are excited that will lead to erosion of planetesimals 
for sizes $ \le 10$km which we have considered here, and growth is
uncertain for 10km sized bodies.

\subsection{High inclination cases and the Kozai effect}
For large inclinations between the binary and planetesimal orbital planes, 
it is possible that the Kozai effect will switch on, causing 
cyclic variations of planetesimal eccentricities and inclinations
\citep{kozai}.
In this section we describe this effect in more detail in order to 
simplify the understanding of results which are described later in this 
paper.
The equations describing the secular evolution of the inclination $\alpha_i$, 
eccentricity, $e_i$, and longitude of pericentre, $\omega_i$,
can be written as \citep{innanen}:
\begin{eqnarray}
\frac{\partial \alpha_i}{\partial\tau}&=&-\frac{15}{8} 
\frac{e_i^2
\sin{(2\omega_i)} \sin{(\alpha_i}) \cos{(\alpha_i)}}{\sqrt{1-e_i^2}}
\label{I-k}\\
\frac{\partial e_i}{\partial\tau}&=&\frac{15}{8}e_i\sqrt{1-e_i^2}
\sin{(2\omega_i)} \sin^2{(\alpha_i)}
\label{e-k}\\
\frac{\partial\omega_i}{\partial\tau}&=&\frac{3}{4}\sqrt{1-e_i^2}\left\{2(1-e_i^2)+5
\sin^2{(\omega_i)}\left[ e_i^2-\sin^2{(\alpha_i)}\right] \right\}
\label{omega-k}
\end{eqnarray}
where for simplicity of notation we have introduced the time variable:
\begin{eqnarray}
\tau=[D^3\Omega_{Ki}/GM_B]^{-1}t.
\label{tau_kozai}
\end{eqnarray}
Hence the time scale on which any of the given quantities change is
$\propto D^3$.
It can be seen from Eq.~(\ref{omega-k}) that there is a value of
$\omega_i$ for which $\dot{\omega}_i=0$.  To lowest order in $e_i$ this occurs when
\begin{eqnarray}
\sin^2{(\omega_i)}=\frac{2}{5 \sin^2{(\alpha_{i0})}}
\label{condition}
\end{eqnarray}
where $\alpha_{i0}$ is the initial inclination of planetesimal $i$.
Having obtained this value, the apsidal precession is halted, as $\dot{\omega}_i=0$.
This leads to the exponential growth of eccentricity. To illustrate this we can 
substitute Eq.~(\ref{condition}) into Eq.~(\ref{e-k}) to find:
\begin{eqnarray}
\frac{\partial e_i}{\partial\tau}&=&\frac{15}{4}\sin^2{(\alpha_{i0})}
\sin{(\omega_i)} \cos{(\omega_i)}e_i \nonumber\\
&=&\frac{15}{4}e_i\sqrt{\frac{2}{5}\left[\sin^2{(\alpha_{i0})}-\frac{2}{5}\right]}.
\label{kozaitime}
\end{eqnarray}
Hence the Kozai effect is only active for initial inclinations 
$\sin{(\alpha_{i0})} \ge \sqrt{\frac{2}{5}}$ or written differently
$\alpha_{i0}\ge 0.68$ radians. The critical angle for which the Kozai
effect switches on is thus $\alpha_K= 39.2^\circ$.
Additionally it can be shown that the Delaunay variable, $D_i$, (which is
equivalent to the component of the planetesimal angular momentum 
which is parallel to the binary angular momentum vector) is a constant
of the motion, where $D_i$ is defined by
\begin{eqnarray}
D_i=\sqrt{a_i(1-e_i^2)} \cos{(\alpha_i)}.
\label{zang}
\end{eqnarray}
As secular variations do not change the semi-major axes, 
this implies that an increase in eccentricity is coupled to a 
decrease in inclination, and vice versa.
\begin{figure}[!b]
\begin{center}
\includegraphics[width=80mm,angle=0]{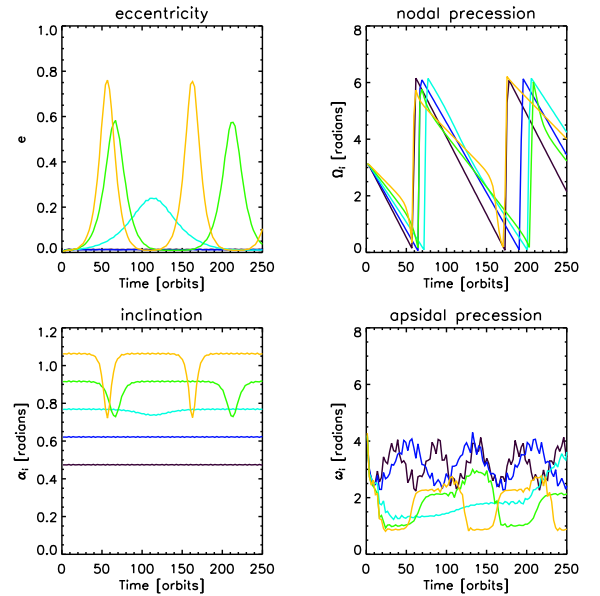}
\end{center}
\caption{Orbital elements of planetesimals with different initial inclinations 
interacting with the binary system only.
Panel 1 shows the eccentricity and panel 2 shows the nodal precession angle.
The inclination with respect to the binary plane and the longitude of 
pericentre (apsidal precession) are depicted in panels 3 and 4, respectively.}
\label{kozai2}
\end{figure}

To study the Kozai effect in the absence of the disc, we have
performed N-body simulations of planetesimals interacting with the binary system only.
The results are depicted in Fig.\ref{kozai2} for planetesimals with various initial 
inclinations. It can be seen that the eccentricities and inclinations undergo 
cyclic variations for planetesimals whose initial inclination are
$\alpha_i\ge\alpha_K$ (yellow, green and light blue lines in Fig.\ref{kozai2}).
This can be explained as follows.
As the eccentricity grows and the inclination decreases (panels 1 and 3), the
latter eventually hits the Kozai threshold, $\alpha_K$, at which point 
there is no value for $\omega_i$ for which expression 
(\ref{condition}) holds. Consequently $\dot{\omega}_i\neq 0$ and $\omega_i$ 
evolves to values for which $\dot{e}_i\le 0$ and $\dot{\alpha}_i\ge 0$. Hence,
eventually $\dot{\omega}_i=0$ can be obtained again, but now for a value of 
$\omega_i$ which is causing the opposite evolution of eccentricities and 
inclinations until the original configuration is recovered
and the Kozai-cycle is completed. We can clearly see in Fig.\ref{kozai2} 
how the increase or decrease of eccentricities and inclinations is connected 
to the halting of apsidal precession during each half-cycle of the 
Kozai mechanism (panel 4). Additionally, we observe that the time scale 
on which the eccentricities and inclinations vary are increased as the initial 
inclination $\alpha_{i0}$ is decreased.
Once the initial inclination is $\alpha_{i0}<\alpha_K$ 
(Fig.\ref{kozai2}, blue and black lines) the condition 
$\dot{\omega_i}=0$ can never be obtained. Consequently there are no net 
changes in eccentricities and inclinations and the Kozai effect is switched off.

\subsection{High inclination case (model 3)}
In model 3 we increased the inclination between the disc-planetesimal system
and the binary plane to
$\gamma_F=0.78$ ($45^\circ$).
As described in the previous section, once the inclination exceeds a critical value 
$\alpha_K=0.68$ ($39.2^\circ$), the Kozai effect may lead to large 
eccentricity and inclination variations of the planetesimals, provided
the disc mass is not too large (see later for a discussion about the
conditions under which the Kozai effect operates).
\begin{figure}[!b]
\begin{center}
\includegraphics[width=80mm,angle=0]{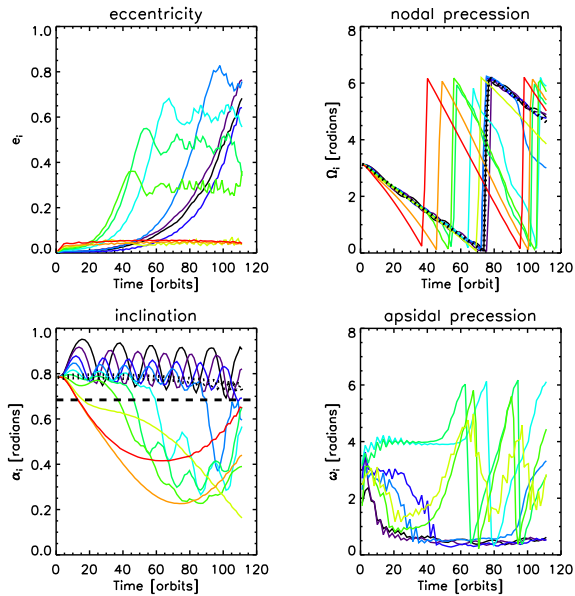}
\end{center}
\caption{Orbital elements for the highly inclined case 
$\gamma_F=0.78 \; (45^\circ )$ (model 3). Panel 1 shows the eccentricity grows 
due to the Kozai effect for all but the outermost planetesimals . The
nodal precession and inclination angles are depicted in panel 2 and 3,
respectively, where the short dashed lines
represent the inner and outer disc edge. The long dashed line in panel 
3 represents the threshold inclination
of $\alpha_K=39.2^\circ$ above which the Kozai mechanism is expected to operate.
The apsidal precession
angles are depicted in panel 4.} \label{highinc}
\end{figure}
The presence of our scaled minimum mass solar nebula disc does not
prevent the onset of the Kozai effect for
planetesimals with semi-major axes $a<12$ AU,
as seen from Fig.\ref{highinc} (panels 1 and 3), where we use the same 
colour convention for depicting the planetesimals as a function
of initial semi-major axis as in models 1 and 2.

The eccentricity grows earliest for planetesimals with larger semi-major
axes (panel 1, green lines).
When the Kozai mechanism starts operating, the inclination starts to decrease. 
In the absence of the disc, this
decrease would continue until the threshold of 
$\alpha_K=0.68 \; (39.2^\circ )$ is approached, after which the inclination
would increase and the eccentricity would decrease again. In the presence of the disc,
however, the situation is
different. As the Kozai effect reduces the inclination, $\alpha_i$
the nodal precession induced by the binary accelerates according to Eq.~(\ref{free}).
This causes the planetesimal orbits to become significantly
inclined relative to the disc midplane (see panel 3 of Fig.\ref{highinc}). 
Concurrent precession about the disc angular momentum vector 
due to the disc gravity causes a 
quasi-monotonic decrease of inclination relative to the binary.
The planetesimals are thus perturbed by the disc onto orbits
with inclination $\alpha_i<\alpha_K$, causing the Kozai effect to switch off. 
At this stage the planetesimal has only
completed a portion of its Kozai cycle and is left with high eccentricity (panel 1)
and a reduced inclination (panel 3), at least for the duration of the simulation.
The apsidal precession angles are depicted in panel 4 of Fig.\ref{highinc}. 
During the time when the Kozai effect is
operating ($\alpha_i>\alpha_K$), the apsidal precession is halted such that 
$\dot{\omega}_i=0$. After
the Kozai effect has switched off ($\alpha_i<\alpha_K$) we observe 
prograde apsidal precession
($\dot{\omega_i}>0$). Although this precession rate is dominated by the perturber 
the disc enhances the prograde
precession rate, since $\delta_i>\alpha_d$ at this stage. 
The total rate is thus given by the sum of contributions by the
binary companion and disc.

Planetesimals with semi-major axes $a_i>12$ AU decouple from the disc due to 
the differential precession
induced by the binary companion. Consequently their inclinations get perturbed 
below $\alpha_K$ before the
Kozai effect can start to operate, and their eccentricities stay low,
at least for the duration of the simulations that we present here. 

The simulation presented in Fig.\ref{highinc} was only run for
110 orbits measured at the disc outer edge ($\sim 10^4$ years).
Those planetesimals which have experienced the Kozai effect
have only undergone half a Kozai cycle, which stalls because the
inclination falls below $\alpha_K$. But we see in panel 3 that the 
inclination relative to the binary is increasing toward $\alpha_K$
at the point when the simulation ends, because the planetesimals
have precessed nodally by more than $180^\circ$ with respect to the disc.
We would therefore expect that the Kozai effect will switch on again
when $\alpha_i > \alpha_K$, allowing the previous Kozai cycle to complete.
Similarly, we see that the inclinations of the planetesimals located 
beyond 12 AU are also increasing toward $\alpha_K$, such that the
Kozai effect will switch on for them eventually.
A long term effect of the disc thus appears to be to prolong the period
associated with the Kozai cycles because of its effect on the planetesimal
inclinations. An important consequence of this is that it increases the
dephasing of the Kozai cycles at different locations in the disc, ensuring
that over long times planetesimals with different semi-major axes are
at very different phases of their Kozai cycles. There will therefore be
very large variations in both eccentricity and inclination within the 
planetesimal swarm, leading to very large collisional velocities
between planetesimals that are well separated in their semi-major axes.

\subsubsection{Varying planetesimal sizes (model 3)}
\begin{figure}[!b]
\begin{center}
\includegraphics[width=80mm,angle=0]{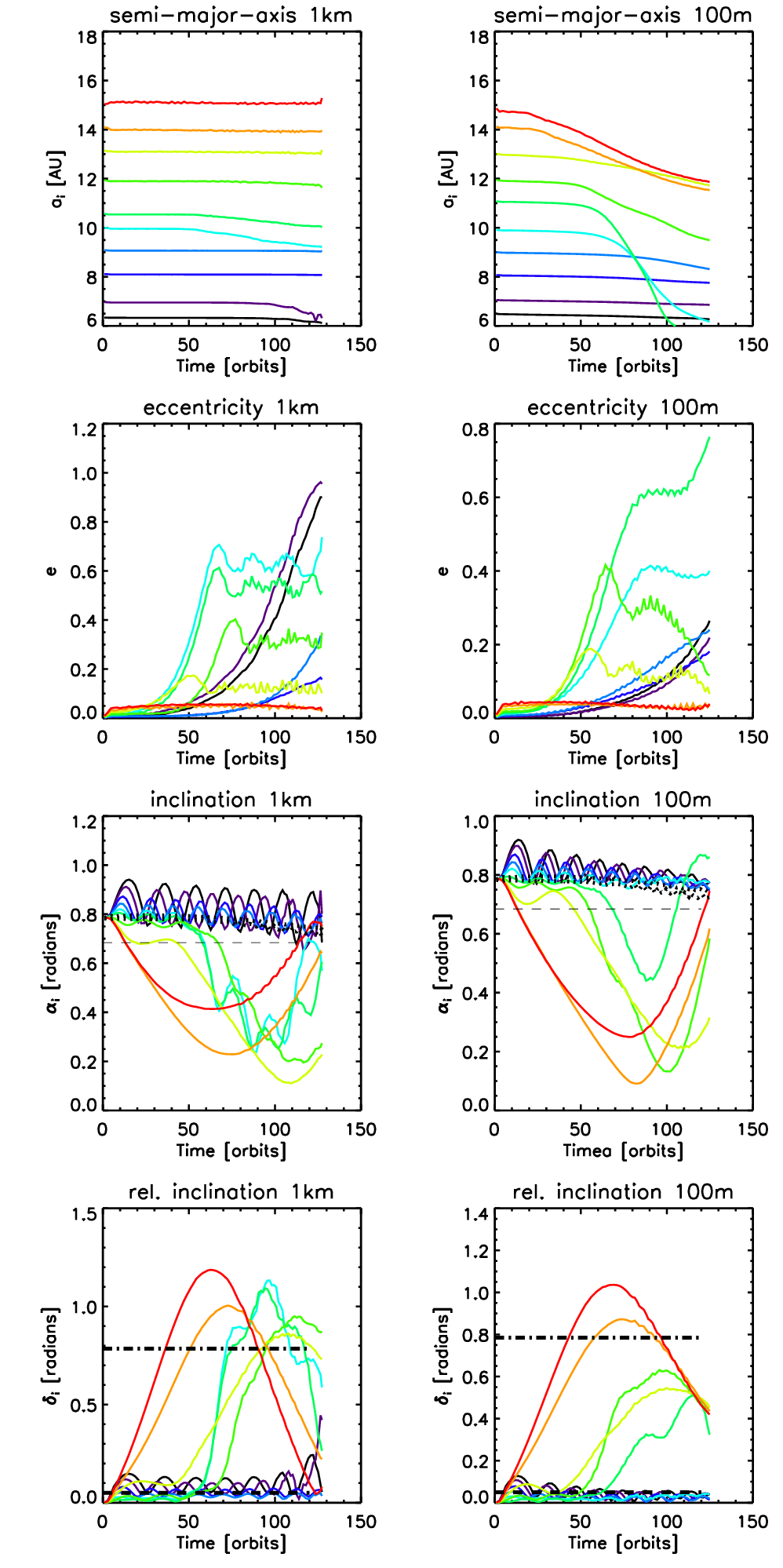}
\end{center}
\caption{Orbital elements for the 1km (left panels) and 100m sized planetesimals 
(right panels). The
semi-major axes are shown in panels 1 and 2. In panels 3 and  4 eccentricity 
growth due to the Kozai effect is still
apparent for both planetesimal sizes. The inclination is shown in panels 5 and 6,
where the short dashed
lines represent the disc inner and outer edge, and the long dashed line marks the 
threshold inclination above which the
Kozai effect operates. In panels 7 and 8 the relative inclination with respect to 
the disc is shown, where the short dashed
line indicates the one pressure scale height limit for the gas disc.
The dash-dotted line represents the inclination of the binary orbit
plane relative to the disc midplane.} \label{sizes}
\end{figure}

In the previous discussion we focused on planetesimals with size
$s_i=10$km, for which the gas drag forces are very weak.
As we decrease the size of the bodies in the system, the gas drag
becomes more important since the stopping time is $\propto s_i$. 
Results for the 1km and 100m sized planetesimals
are shown in the left and right panels of Fig.\ref{sizes} respectively.
The Kozai effect leads 
to the growth of eccentricities (panels 3 and 4) and
relative inclinations (panels 7 and 8) for both particle sizes shown. 
This is not surprising, since the time scale
on which we expect substantial damping of eccentricity and inclination to 
occur is of the order of the gas drag
stopping time. For our disc model this is given by:
\begin{eqnarray}
\tau_S=0.81\left(\frac{s_i}{{\rm m}}\right)\left(\frac{a}{\rm AU}\right)^2 \; P_d,
\label{stoptime2}
\end{eqnarray}
where $s_i$ is expressed in metres.
For the $100$m sized planetesimals at $a=6-15$ AU the stopping time is of the order
of $10^3-10^4$ orbits, respectively, and is a factor of $10$ longer for the 
$1$km sized planetesimals.
This is much longer than the time scale on which the Kozai effect operates 
($\sim 10^2$ orbits),
and hence we would not expect the gas drag to prevent growth of eccentricity 
and relative inclination in this size range.
We comment here (without showing results), that 
we also found the Kozai effect to
operate for $10$m sized bodies, for which the drag and Kozai time scales
are similar.

The increased effect of gas drag causes the semi-major axes to decay 
(see panels 1 and 2 in Fig.\ref{sizes}).
As planetesimals experience the Kozai effect,
and develop large eccentricities and relative
inclinations, their relative velocities with respect to the gas disc
becomes very large (since $|{\bf v}_i-{\bf
v}|^2\simeq v_K^2\left(e_i^2+\delta_i^2\right)$ from expression [\ref{lissauer}]). 
As they travel through the disc they lose kinetic energy rapidly
to the gas, and their semi-major axes decay. We note that those 100m sized
bodies lying beyond 12 AU, which do not experience the Kozai effect during the
simulation, also undergo rapid decay in their semi-major axes.
These particles develop large inclinations relative to the disc
due to the rapid nodal precession induced by the companion, and
the resulting gas drag as they pass through the disc leads
to their orbital decay.

Before the Kozai effect starts to operate we can see that the increased effect of gas 
drag for the 100m sized bodies (right panels of Fig.~\ref{sizes}) causes damping of the 
relative inclination (panel 8) and therefore a decrease of the amplitude 
of the oscillation about the disc midplane caused by disc gravity (panel 6). 
In other words, the increased gas drag causes the 100m sized bodies to remain closer
to the disc midplane, with the oscillations in the amplitude of relative
inclination being damped to a larger degree than occurs for larger bodies. 
This effect will become important when discussing relative velocities
between the differently sized planetesimals, 
as the size dependence of the inclination perturbations will have 
an effect on the Kozai effect operating on the planetesimals, 
causing planetesimals of different sizes to orbit in planes
which are significantly inclined with respect to one another.
We note that the relative inclinations of both the 1km and
10km sized bodies are found to be similar, since the gas drag is weak in
both of these cases.

\subsubsection{Collisional velocities for the highly inclined  case (model 9)}
\begin{figure}[!b]
\begin{center}
\includegraphics[width=80mm,angle=0]{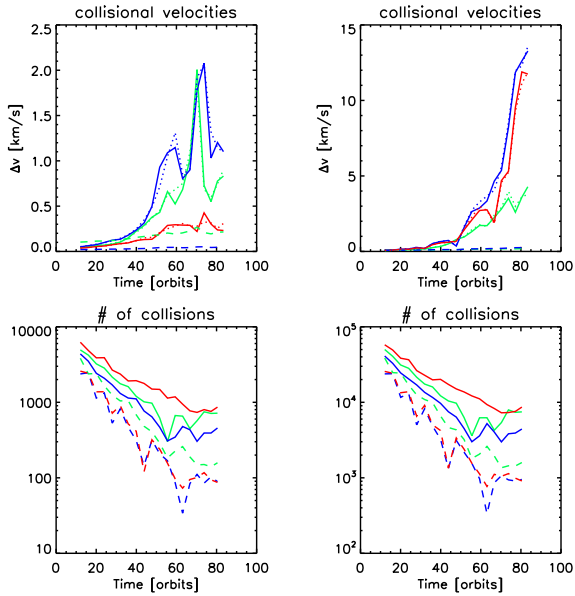}
\end{center}
\caption{Average collisional velocities in the high inclination 
$\gamma_F=0.78 (45^\circ)$ case between equally (panel
1) and differently sized planetesimals (panel 2). Left panel: 10km-10km collisions
(green-solid); 1km-1km collisions (blue-solid);
100m-100m collisions (red-solid). Right panel: 10km-1km (green-solid); 
10km-100m (blue-solid); 1km-100m (red-solid).
Threshold velocities for catastrophic disruption corresponding to the 
different size-combinations for the strong
aggregates (dashed line) are also shown. Relative velocities have also been 
calculated using a larger orbit
width $\Delta a=2\cdot 10^{-3}$ AU (dotted lines). Panels 3 and 4: 
collision count for orbital width $\Delta
a=2\cdot 10^{-4}$ AU and $\Delta a=2\cdot 10^{-3}$ AU, respectively. 
The line colours and styles in panel 3 and 4 are
as follows: green-solid (10km-10km); blue-solid (1km-1km); red-solid (100m-100m);
green-dashed (10km-1km);
blue-dashed (10km-100m); red-dashed (1km-100m).}
\label{vel_highinc}
\end{figure}

In this section we investigate the relative velocities for the high 
inclination ($\gamma_F=0.78$) ($45^\circ$) case.
The procedure to detect collisions and measure relative velocities is 
identical to the one applied to the low
inclination case (model 8). In Fig.\ref{vel_highinc} (panel 1 and 2, solid lines) 
we present collisional velocities between
equally and differently sized planetesimals, respectively.  
Collisional velocities are even higher in this case ($\sim$ a few km/s) 
than for the low inclination case for all size combinations considered.

\begin{figure}
\begin{center}
\includegraphics[width=80mm,angle=0]{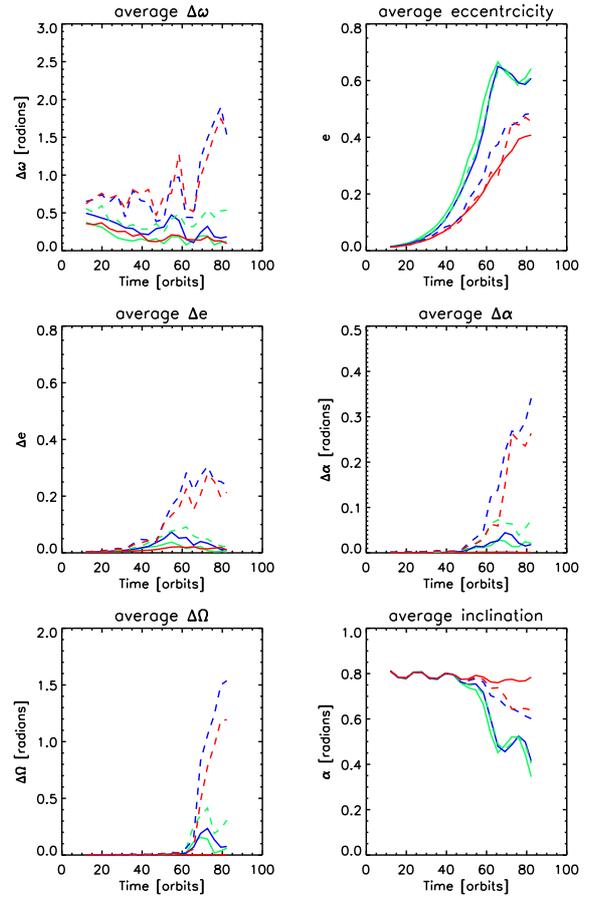}
\end{center}
\caption{Average orbital parameters of the colliding planetesimal pairs 
in the high inclination
$\gamma_F=45^\circ$ case. The line colours and styles are as follows: 
green-solid (10km-10km); blue-solid
(1km-1km); red-solid (100m-100m); green-dashed (10km-1km);
blue-dashed (10km-100m); red-dashed (1km-100m).}
\label{orbital_highinc}
\end{figure}
The orbital elements are displayed in Fig.\ref{orbital_highinc}.
We observe that the Kozai effect causes an increase of eccentricity up to $e=0.4$-0.6,
such that the contribution to the relative velocity between planetesimals
due to the term $\propto e_i\Delta\omega_i$ becomes large,
and this is the dominant cause of high velocity collisions between equal sized bodies.
The situation is different for planetesimals of different sizes.
The different gas drag strengths 
change the relative inclination between the planetesimals and the disc,
resulting in a variation of the inclination perturbations caused by disc gravity, 
which on the one hand causes the value of $\omega_i$ that is approached during the 
first half of the Kozai cycle to be different, and on the other hand causes 
the Kozai mechanism to operate on different time scales.
As a result planetesimals of different sizes decouple from the disc at different 
times.
As can be seen from panels 4 and 5 of Fig.\ref{orbital_highinc},
the orbital planes and pericentres of different sized planetesimals become 
randomly distributed with $\Delta\Omega_i\sim 1$ and $\Delta\omega_i\sim 1$,
and the relative velocities are increased substantially due 
to the terms which are proportional to
$e_i\Delta\omega_i$, $\Delta \alpha_i$, 
and $\sin{(\alpha_i)}\Delta\Omega_i$ in Eq.~(\ref{ana_vel}),
with the latter term being the dominant one.
As found in the low inclination case, the relative velocities of different sized
planetesimals are generated due to misalignment of their pericentres 
$\Delta\omega_i$ and their orbital planes
$\Delta\Omega_i$. In contrast to the low inclination model, however,
the eccentricities, $e_i$, inclinations $\alpha_i$,
and the differential nodal precession angle,
$\Delta\Omega_i$, are much larger due to the
Kozai effect, leading to very large collisional velocities.

In Fig.\ref{vel_highinc} (panel 3) we display the number of collisions detected
for the different size combinations, and it may be observed that
the number of collisions drops to $\sim 100$ for collisions
between differently sized planetesimals. As the bodies begin to orbit in 
different planes their encounter
probability is reduced, as discussed in Sect.~\ref{sec:low-inc-coll} 
and reported by \cite{xie}. 
This implies that the averaged relative velocities
(panels 1 and 2) are obtained from only $\sim 10^2-10^3$ data points, and 
the results are in danger of becoming 
statistically unreliable. In order to compare these collision
velocities with more statistically significant data,
we also calculated the relative velocities while adopting a
larger cross sectional area for the intersecting orbits corresponding to 
$\Delta r=2\cdot 10^{-3}$ AU. The collision
probability is increased by a factor of $\sim 10$,
as observed in panel 4 of Fig.\ref{vel_highinc}.
These relative velocities are plotted as dotted lines in panels 1 and 2
of Fig.~\ref{vel_highinc},
and are almost indistinguishable from the data
obtained with the smaller orbit width, suggesting that the results are reliable.
As was discussed for the low inclination model, the reduction
in the collision frequency can have potentially important consequences for
planetesimal accretion, since it may favour collisions between similar sized
bodies which orbit in similar planes. But, it is clear from Fig.\ref{vel_highinc}
that when the Kozai effect is active, large collision velocities are
obtained for all size combinations.

The disruption/erosion threshold velocity for the strong aggregates is shown using
the dashed lines in panels 1 and 2. The
collisional velocities are substantially larger than those required for
erosion or catastrophic disruption for all size combinations in this case, leading
to the unsurprising conclusion that
planetesimal accretion is not possible if the Kozai mechanism operates
during the epoch of planet formation. For the relative
velocities observed in Fig.\ref{vel_highinc} of $\sim 2$km/s for 
equally sized planetesimals, we estimate that collisions will always lead to
fragmentation unless bodies of size $\sim 10^3$ km have already formed.

\section{Suppressing the Kozai effect}

\subsection{Increasing the disc mass (models 4 and 5)}
The Kozai effect generates relative velocities which are 
always too large for collisional growth of planetesimals, at least for
sizes in the range (100m-10km).
An interesting question is under what conditions does the Kozai mechanism 
cease to operate.
As discussed in Sect. 5.3 the Kozai mechanism relies on the
apsidal precession induced by the binary companion halting, 
such that $\dot{\omega}_i=0$, which then allows for a secular 
change of the eccentricity and inclination values.
If apsidal precession can be induced by another source, however,
such as the disk for example, then we would expect the Kozai mechanism to
be ineffective once this induced precession is fast enough \citep{wu}.

\begin{figure}[!t]
\begin{center}
\includegraphics[width=80mm,angle=0]{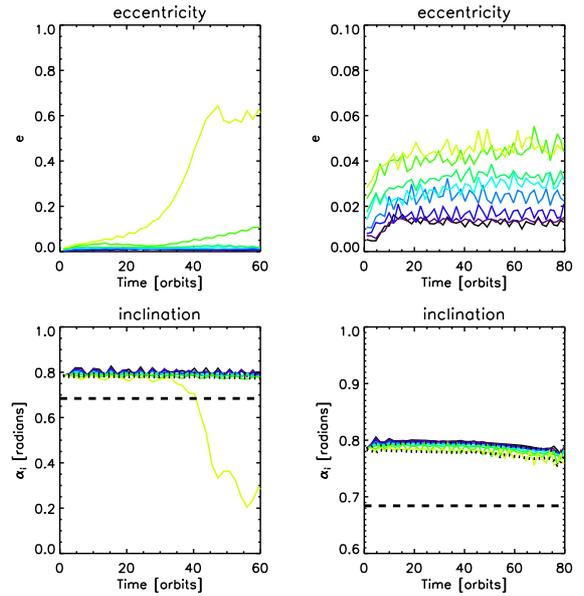}
\end{center}
\caption{Orbital elements for model 4 (left panels) and model 5 (right panels). 
Only planetesimals with
semi-major axes $a\le 13$ AU are included. In model 4 the Kozai effect is still 
operating, while in model 5 it is not.
Panels 1 and 2 show the eccentricity. Panels 3 and 4 display the inclination,
where the short dashed
lines represent the inner and outer edge of the disc and the long dashed line 
indicates the threshold
inclination of $0.68 \; (39.2^\circ )$ above which the Kozai mechanism can operate.} 
\label{mass}
\end{figure}

Since the apsidal precession
rate caused by the disc scales with disc mass (see Eq.\ref{disc_aps}), 
we have run simulations with an increased disc mass to see if the
Kozai mechanism can be suppressed in this way. The results are shown in 
Fig.\ref{mass} for 
model 4 (left panels) and 5 (right panels) with disc masses of 
$M_d=3.3\hat{M}$ and $6.6\hat{M}$, respectively, where $\hat{M}$ is the
nominal disc mass used in model 3.
As can be seen from the figure, the Kozai mechanism still switches on for 
the $M_d=3.3\hat{M}$ case
but does not operate in the $M_d=6.6\hat{M}$ model. Consequently 
the eccentricity grows to large values in model 4, as seen in panel 1
of Fig.\ref{mass}, while it remains low in model 5 (panel 2).

\subsection{Increasing binary separation (models 6 and 7)}
Another way to increase the relative importance of the disc-induced apsidal 
precession is to decrease the gravitational influence of the binary companion.
In this section we present simulations in which the separation of the binary system 
was increased to examine when the Kozai effect is suppressed.
We consider separations of $D=90$ AU and $D=120$ AU in models 6 and 7, respectively.
Because the Kozai time scale is $\propto D^3$, extremely long simulation 
times are required to prove the existence
(or non existence) of the Kozai effect.
The required run times would be prohibitive for a full 3D hydrodynamic
simulation, so we adopted the compromise of
switching off the hydrodynamic evolution of the disc in order 
to speed up the run times for these models. The disc in these simulations
therefore remains axisymmetric and static in the precessing frame, but 
our adoption of a precessing reference frame causes it to precess
around the binary angular momentum vector at the prescribed rate. 

We tested the accuracy of this
approach by re-running model 3 using a static, precessing disc model, and found
similar results when compared with the run in which the disc was allowed
to evolve self-consistently. Differences between the static, precessing disc model
and the full hydrodynamic simulation arise mainly because the disc
experiences a low amplitude nodding motion (oscillation in its inclination) with a period
equal to half the binary orbit period in the full simulation \citep{larwood, fragner},
where this is driven by the time-varying gravitational field of the binary.
Although the effect is relatively small, it is likely to reduce the accuracy
with which a rigidly precessing disc model can be used to determine planetesimal
collision velocities in detail. But, it is a useful set-up for determining
the disc mass for which the Kozai mechanism operates.

\begin{figure}[!t]
\begin{center}
\includegraphics[width=80mm,angle=0]{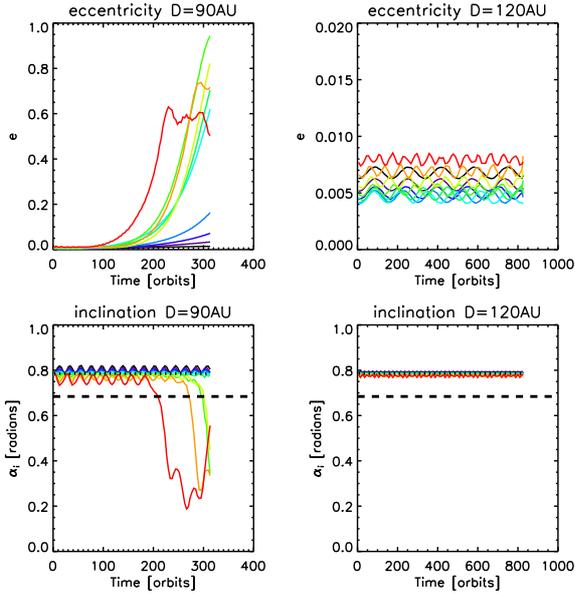}
\end{center}
\caption{Orbital elements for model 6 (left panels) and model 7 (right panels). 
The Kozai effect is operating
in model 6, but is not in model 7. The eccentricities are shown in panels 1 and 2. 
The inclination is shown in panels 3 and 4. The long dashed lines in panels 3
and 4 represents the $39.2^\circ$ inclination threshold,
above which the Kozai effect can operate. } \label{separations}
\end{figure}

The results are shown in Fig.\ref{separations} for the 
$D=90$ AU (left panels) and $D=120$ AU model (right panels).
We can observe that the Kozai effect is operating in model 6 while it 
is not in model 7.
The time scale for the Kozai effect to increase the eccentricities is 
larger by a factor $3.3$, as expected, for model 6
compared to the reference model 3 due to the Kozai time scale increasing as
$\sim D^3$.
If the Kozai effect was operating in model 7, we would expect to see substantial
eccentricity growth after 500 orbits (a factor 8 longer than in our reference model). 
This is not seen, however,
and we conclude that the Kozai effect cannot operate in model 7.

\subsection{A theoretical argument}
\begin{figure}[!t]
\begin{center}
\includegraphics[width=80mm,angle=0]{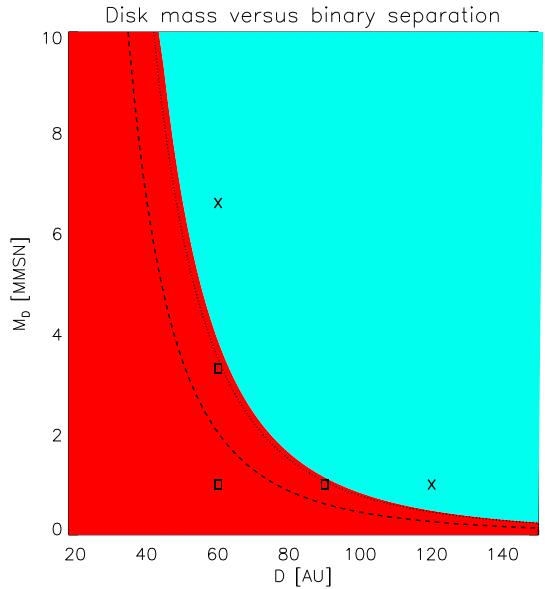}
\end{center}
\caption{Plot of the parameter regime explored in the simulations. 
The diagram shows disc mass against binary separation, with
the red area denoting the region where the Kozai mechanism should
operate, and the cyan area showing the region where it should not
for a planetesimal located at 11 AU.
The symbols denote the
outcome of the simulations, where a cross implies that the Kozai effect
was switched off, and an open square indicates that the Kozai mechanism
operates. The dashed and dotted line represents the
same boundary for a body at $a_i=6$ AU and $a_i=15$ AU respectively.} \label{kozai}
\end{figure}

We have seen in the previous two sections that the Kozai mechanism can be 
suppressed provided the disc mass or the
binary separation are large enough. Now we want to understand the
numerical results by means of a theoretical argument.
As has already been discussed, the Kozai mechanism relies on halting the 
apsidal precession of the planetesimals induced by the binary companion. 
If this can be prevented then there should be no secular net change of 
eccentricities or inclinations.
The total net rate of apsidal precession is given by the sum of the 
contributions by the binary companion (Eq.\ref{omega-k}) and the disc induced 
apsidal precession. Since we only consider cases for 
which $\delta_i\le\alpha_d$, the latter is given by Eq.~(\ref{disc_aps}).
Hence the total apsidal precession rate experienced by a planetesimal on 
a circular orbit ($e_i=0$) is:
\begin{eqnarray}
\frac{\partial\omega_i}{\partial t}=\frac{3 \pi}{2 D^3} \left(20 a_i
\right)^\frac{3}{2} \left(2-5 \sin^2{(\omega_i)} \sin^2{(\alpha_i)}\right)-|\dot{\omega}_D|,
\label{domegadt}
\end{eqnarray}
where $a$ and $D$ are expressed in units of AU.
The first term on the right hand side of 
Eq.~(\ref{domegadt}) accounts for the binary-induced 
precession, and $\dot{\omega}_D$ is the disc-induced precession rate
given by Eq.~(\ref{disc_aps}). Both of these rates are expressed in time units of 
orbits at $a=20$ AU (the time unit used throughout the paper). 
In order for the Kozai mechanism to
operate we require that $\frac{\partial\omega_i}{\partial t}=0$. This can only 
be true if the prograde
term due to the binary companion is larger than the retrograde term of the disk.
For the Kozai effect to operate, we therefore require that:
\begin{eqnarray}
\frac{3\pi}{D^3} \left(20 a_i \right)^\frac{3}{2}\ge|\dot{\omega}_D|.
\end{eqnarray}
Using Eq.~(\ref{disc_aps}), we can express this a condition on the total
disc mass in terms of the planetesimal semi-major axis and binary orbit separation:
\begin{eqnarray}
M_d \le \frac{3\pi\hat{M}}{D^3}\left( 20 a_i \right)^\frac{3}{2}
\left[C_0+C_1\left(\frac{a_i}{{\rm AU}}\right)+C_2
\left(\frac{a}{{\rm AU}}\right)^2\right]^{-1}.
\end{eqnarray}
If this inequality is fulfilled we expect the Kozai mechanism to operate. 
In Fig.\ref{kozai} we show how the upper limit on the
disc mass varies as a function of binary separation in order for the Kozai
mechanism to just switch on/off for a planetesimal orbiting at 
a representative value of the semi-major axis $a_i=11$ AU, which is 
approximately half of the disc tidal truncation radius.
The red area corresponds to the regime
where the inequality is fulfilled and the Kozai-mechanism should operate. 
The cyan area marks the
parameter space where the Kozai mechanism should not operate
(large disc mass/large binary separation).
The symbols represent the simulation results, where open squares indicate that the Kozai
mechanism was found to operate (models 3, 4 and  6), while crosses represent the 
models in which the Kozai effect was
ineffective (models 5 and 7). We see that the numerical results agree with our
predictions. 

We also plot the boundary lines that correspond to semi-major 
axes of $a_i=6$ AU (dashed line) and $a_i=15$ AU
(dotted line). Between $a_i=6$ AU and $a_i=11$ AU the precession frequency due 
to the disc is nearly independent of
semi-major axis, as can be seen from the solid line in panel 7 of
Fig.\ref{od}. Yet the binary precession still
increases as $\propto a^\frac{3}{2}$. Hence the parameter space for which the 
Kozai mechanism operates becomes
larger as the semi-major axis increases. Between $a_i=11$ AU and $a_i=15$ AU 
the disc precession rate increases in
magnitude approximately as $\propto a_i^{3/2}$. Thus the boundary separating 
the Kozai-active versus Kozai-inactive regions does not
change very much beyond $a_i=11$ AU.

\section{Discussion and conclusions}
In this paper we have investigated the dynamics of planetesimals that
are embedded in a gaseous disc that is perturbed by a binary
companion on a circular, inclined orbit. 
In contrast to previous work the planetesimals are allowed to interact with the
gaseous disc via gas drag {\it and} disc gravity. 
The disc evolves hydrodynamically
because of gravitational perturbations due to
the binary companion, and the disc model we consider 
undergoes solid body precession around the
orbital angular momentum vector of the binary system,
as expected when the warp propagation time {\it via}
bending waves is shorter than the differential precession time.
Discs around young stars are expected to show similar behaviour
since it is estimated that $\alpha < H/R$ in these systems,
where $\alpha$ is the usual viscosity parameter.
The key results of our study can be summarized as follows.

\begin{itemize}

\item In the absence of disc gravity, planetesimals undergo strong
differential nodal precession, such that they eventually
orbit in different planes \citep{marzari2, xie}. They also precess relative to the
disc, so their orbits become highly inclined relative to it.
We find that disc gravity acts to
prevent this differential precession, and forces the planetesimals
to precess with the disc on average. Viewed locally,
the forcing of nodal precession by the binary companion causes the   
planetesimals to oscillate about the midplane of the disc, with an
amplitude that depends on the size of the planetesimals because of
the influence of gas drag in damping this relative motion. This key result
suggests that the influence of the gravitational field of the precessing
disc should always be included in future studies of planet formation in
inclined binary systems.

\item Previous studies that focused on the influence
of a coplanar, eccentric binary companion have shown that
the eccentric binary generates a forced eccentricity 
in the planetesimal swarm. Gas drag acts to align the
pericentres of planetesimals which are of the same size, such
that collisions between same-sized bodies are dominated
by keplerian shear. Collisions between different sized
bodies occur with higher velocity due to the misaligned
orbits \citep{marzari, thebault}. We find that for a
circular binary, where the eccentricity of planetesimals
is largely generated by high frequency terms in the disturbing function,
strong alignment of pericentres is not observed. 

\item Previous work which has examined planetesimal
dynamics in modestly inclined and eccentric binary systems suggests
that gas drag causes size-dependent phasing of both pericentres
and the lines of nodes of perturbed planetesimals \citep{xie}.
This has the effect of favouring collisions between same sized objects,
for which the collision velocity is dominated by the keplerian shear,
and as such it has been suggested that this may 
provide an effective channel for planetesimal growth in inclined
binary systems. Our results obtained for a binary with
inclination $\gamma_F=25^\circ$ are in basic agreement with this,
as we find that same-sized planetesimals do indeed occupy very similar
orbital planes, whereas the orbits of differently sized bodies are
mutually inclined. This arises in our case because of the different
amplitudes of oscillation about the disc midplane observed for
planetesimals of different size, with smaller bodies remaining
closer to the midplane. We thus find that the collision
velocities between differently sized bodies are significantly
larger than between same-sized bodies 
(typically $v_{coll} \simeq 200$ ms$^{-1}$ for different sized objects and
$v_{coll} \simeq 50$ - 70 ms$^{-1}$ for same-sized objects).
We note that the lack of pericentre alignment observed
in our simulations leads to the substantial collision velocities between
same size bodies. Collisions with the velocities described above are likely to
be erosive or disruptive, depending on the mass ratio of the colliding bodies.

\item
For highly inclined systems with $\gamma_F = 45^\circ$,
we find that the Kozai mechanism can operate,
causing large changes in the eccentricities and
inclinations of the planetesimals, and leading to
collisional velocities that are much too large to
allow for planetesimal accretion. The long term influence of
the disc in cases where the Kozai mechanism operates
is to modify the period associated with the Kozai cycle
by periodically forcing the planetesimal inclinations to
fall below, or rise above, the critical value for the
Kozai effect to operate, $\alpha_K=39.2^\circ$.
We find that planetesimals of size 10m - 10km 
all experience the Kozai effect.

\item
Increasing the disc mass or binary separation can suppress the
Kozai effect, as disc gravity can induce apsidal precession 
which is fast enough to render the Kozai mechanism ineffective.
For a binary separation of $D=60$ AU, and disc truncation radius
of 20 AU, we find that we need to increase the disc mass by
a factor of $\sim 6$ above the minimum mass solar nebula (MMSN) value in
order to switch off the Kozai effect at a representative orbital
radius of $a_i=11$ AU. Increasing the binary separation to 
$D=120$ AU, and maintaining the mass of our disc at its
nominal value (equivalent to the MMSN) also rendered the Kozai
mechanism inoperative.

\end{itemize}

In the light of the above findings, we can conclude that
highly inclined, distant binary companions probably will not
induce the Kozai effect during planet formation while the
disc is present, although it may do so once the disc has
dissipated, such that the final planetary system is 
strongly perturbed. But we note that the peak in the distribution of
binary orbit separations occurs at $D \simeq 30$ AU 
\citep{duquennoy, ghez, leinert},
which is small enough for the Kozai mechanism to operate during
the planet formation epoch when the gas disc is present.
There are number of extrasolar
planetary systems, however, which are observed to be in
relatively close binary systems, $\gamma$ Cephei being
a notable example (see Kley \& Nelson 2008, and references therein).
Here the binary separation is only a few tens of AU, such
that the Kozai mechanism could possibly have operated
when planetesimals were accreting. The fact
that a planet is observed in this system with a modest orbital
eccentricity is probably an
indication that the binary inclination is too small to allow
the Kozai mechanism to operate. This is consistent with the
findings of \citet{hale} that binary orbits with $D < 40$ AU
tend to be reasonably well aligned with the stellar spin axes.

A key issue that needs to be addressed in the future is the
evolution of a planetesimal swarm in a binary system where
the orbit is both inclined and eccentric. Our results suggest
that a circular binary system does not allow for the pericentres
of perturbed planetesimals to be well aligned, since a circular
orbit is unable to impose a preferred direction on the system,
and this is in clear contrast to studies which have considered
an eccentric companion. It should be noted, however, that the
inclusion of the disc gravity will be of crucial importance in such
a study, since it will significantly perturb the orbits of the
planetesimals. This is particularly the case when the planetesimal
orbits are inclined relative to the disc midplane, and the 
presence of an eccentric binary companion may also cause the
disc itself to become eccentric \citep{kley,kley-papaloizou}
(where the disc eccentricity may be dependent on the
disc self-gravity \citep{Marzari-edisc}),
further complicating matters. We further note that the inner edge
of our computational domain was located at $R=2$ AU, so our present 
study examines planet formation in the region normally associated
with giant planet formation. Studies of this type which examine
terrestrial planet formation closer to the central star will
require deployment of substantially more powerful computational resources
than were used here, in order to simulate the larger range of time scales 
in the problem. Such a study should also include a more realistic
planetesimal size distribution so that the possibilty of accretion
occuring via collisions between similarly sized bodies can be explored.

In this paper we have only considered planet formation via collisions between
planetesimals, whereas dust accumulation onto individual planetesimals could 
still lead to planetary growth as long as collisions between planetesimals can 
be avoided. This issue has been recently been addressed by 
\cite{paardekooper2, xie2010}, where they show that km-sized planetesimals may grow 
in size by two orders of magnitude due to accretion of collisional
debris if the efficiency of 
planetesimal formation stays low, i.e. only a few planetesimals form. This study
involved two-dimensional simulations, and it will be interesting to examine how
the results change for a misaligned system where planetesimals may orbit
out of the disc midplane where the majority of the dust and collisional
debris will reside. As we have seen in Sect. 5.2, 10 km sized planetesimals 
tend to have orbits which are inclined relative to the disc 
by more that the disc scale height, once their free particle rates are substantially
different from the disc precession rate. Planetesimal formation via dust accretion is 
therefore only expected in radial regions of the disc where the precession rate
of the planetesimals and the disc are well-matched.
Closer to the inner or outer edge of the disc,
efficient dust accretion may not be possible due to the large 
relative inclinations of the planetesimals. This, and other issues discussed in this
paper, will be the subject of future publications.

\acknowledgements
The simulations presented in this paper were performed on the QMUL
HPC facility purchased under the SRIF initiative. We gratefully acknowledge
detailed and insightful comments provided by the referee, Phillipe Th\'ebault,
which led to significant improvements in this paper. MMF gratefully acknowledges
the support of Stephan Rosswog and staff at the University of Bremen,
where part of this work was carried out. 

\bibliographystyle{aa}

\bibliographystyle{aa}

\listofobjects
\end{document}